\documentclass{emulateapj}

\newcommand{\iso}{{\em ISO}}
\newcommand{\iras}{{\em IRAS}}
\newcommand{\mum}{\ifmmode{\rm \mu m}\else{$\mu$m}\fi}

\begin{document}

\title{The Magellanic zoo:  Mid-infrared {\it Spitzer} 
spectroscopy of evolved stars and circumstellar dust in 
the Magellanic Clouds}

\author{
G.~C.~Sloan\altaffilmark{1},
K.~E.~Kraemer\altaffilmark{2},
P.~R.~Wood\altaffilmark{3},
A.~A.~Zijlstra\altaffilmark{4},
J.~Bernard-Salas\altaffilmark{1},
D.~Devost\altaffilmark{1,5},
J.~R.~Houck\altaffilmark{1}
}
\altaffiltext{1}{Cornell University, Astronomy Department,
  Ithaca, NY 14853-6801, sloan@isc.astro.cornell.edu, 
  jbs@isc.astro.cornell.edu, jrh13@cornell.edu}
\altaffiltext{2}{Air Force Research Laboratory, Space Vehicles
   Directorate, 29 Randolph Road, Hanscom AFB, MA 01731}
\altaffiltext{3}{Research School of Astronomy and Astrophysics,
  Australian National University, Cotter Road, Weston Creek ACT 2611,
  Australia, wood@mso.anu.edu.au}
\altaffiltext{4}{Univ. of Manchester, School of Physics \& Astronomy,
   P.~O.~Box 88, Manchester M60 1QD, albert.zijlstra@manchester.ac.uk}
\altaffiltext{5}{Canada France Hawaii Telescope, 65-1238 Mamalahoa 
  Hwy, Kamuela, HI, 96743, devost@cfht.hawaii.edu} 

\begin{abstract}

We observed a sample of evolved stars in the Large and Small 
Magellanic Clouds (LMC and SMC) with the Infrared Spectrograph 
on the {\it Spitzer Space Telescope}.  Comparing samples from
the SMC, LMC, and the Galaxy reveals that the dust-production
rate depends on metallicity for oxygen-rich stars, but carbon
stars with similar pulsation properties produce similar
quantities of dust, regardless of their initial metallicity.
Other properties of the oxygen-rich stars also depend on
metallicity.  As the metallicity decreases, the fraction of 
naked (i.e.\ dust-free) stars increases, and among the naked
stars, the strength of the 8~\mum\ absorption band from SiO 
decreases.  Our sample includes several massive stars 
in the LMC with long pulsation periods which produce 
significant amounts of dust, probably because they are young 
and relatively metal rich.  Little alumina dust is seen in 
circumstellar shells in the SMC and LMC, unlike in Galactic 
samples.  Three oxygen-rich sources also show emission from 
magnesium-rich crystalline silicates.  Many also show an 
emission feature at 14~\mum.  The one S star in our sample 
shows a newly detected emission feature centered at 
13.5~\mum.  At lower metallicity, carbon stars with similar 
amounts of amorphous carbon in their shells have stronger 
absorption from molecular acetylene (C$_2$H$_2$) and weaker 
emission from SiC and MgS dust, as discovered in previous 
studies.  

\end{abstract}


\section{Introduction} 

In the Galaxy, stars on the asymptotic giant branch (AGB)
may be responsible for nearly 90\% of the dust injected into
the interstellar medium (ISM), with supergiants accounting
for most of the remainder \citep{geh89}.  How much these
relative contributions change in the early Universe is not
well understood.  This question has grown even more 
important, now that the Infrared Spectrograph 
\citep[IRS][]{hou04} on the {\it Spitzer Space Telescope} 
\citep{wer04a} has detected dust in galaxies at redshifts 
of 2.6--2.7 \citep{tep07,lut07}, corresponding to look-back 
times of $\sim$11 Gyr in a Universe 13.7 Gyr old.  Supernovae 
have been suggested as a possible source of dust in the early
Universe, \citep[e.g.][]{cmc67,ds80}, but the amount of 
dust they contribute remains controversial 
\citep[e.g.][]{sug06,mei07}.  The most massive stars that can
evolve to the AGB will begin producing dust only $\sim$100 Myr 
after the formation of the first stars in a galaxy.
This initial population will be metal poor.

Nearby galaxies within the Local Group give us the 
opportunity to study how the dust produced by individual
supergiants and AGB stars depends on metallicity.
The Large and Small Magellanic Clouds (LMC and SMC) lie at 
distances of 50 and 61 kpc, respectively 
\citep[e.g.][]{alv04,kw06}.  Measurements of the metallicity 
([Fe/H]) of the LMC vary between $-0.3$ and $-0.4$, while the 
range in the SMC is between $-0.6$ and $-0.8$ \citep[][and 
references therein]{kw06}.

The {\it Infrared Space Observatory} ({\it ISO}) obtained
the first mid-infrared spectra of evolved stars in the LMC.
\cite{tra99} published spectra of 23 targets using the 
PHOT-S spectrometer, which covered the 2--5 and 
6--12~\mum\ wavelength ranges at a resolution $R\sim$90 
($R=\lambda/\Delta\lambda$), and the CAM-CVF 
spectrometer, which observed the 7--14~\mum\ wavelength range 
at a resolution of $\sim$40.  The formation of CO molecules
in stellar outflows will exhaust either C or O from the material
available to form dust and divides the dust chemistry around
these objects into oxygen-rich or carbon-rich.  The \iso\
sample includes spectra from both chemistries, and 
\cite{tra99} showed that they could separate these two
groups photometrically with color-color diagrams.  

The spectrometers on \iso\ allowed observations of only the 
brightest targets.  The improved sensitivity of the IRS on 
{\it Spitzer} has made possible more detailed studies of 
mass-loss and dust injection in both the LMC and SMC.  The 
five major programs have observed a combined total of 176 
sources in the LMC and 63 in the SMC, most of them evolved 
stars.  \cite{slo08a} reviewed these programs and the 
different criteria used to select targets.  

Program 3426 (P.I.\ Kastner) focused on the brightest 
sources in the LMC, using the classes defined from 
observations of the {\it Mid-Course Space Experiment} ({\it 
MSX}) by \cite{msx01} \citep{kas06,buc06}.  Program 3591 
(P.I.\ Kemper) targeted objects covering the stages of 
evolution of oxygen-rich stars from the AGB to planetary 
nebulae \citep{lei08,mk08}.  Program 3277 (P.I.\ Egan) 
selected targets in the SMC based on their infrared colors 
\citep{kra05,kra06,slo06b}.  Program 3505 (P.I.\ Wood) 
spanned a range of mass-loss rates in the LMC and the SMC.  
It included a young stellar object \citep[YSO;][]{vl05b}, 
but nearly all of the rest are carbon stars 
\citep{zij06,mat06,gro07,lag07}.  
The other three programs also observed more carbon stars than 
initially planned, with sources classified photometrically as 
OH/IR stars turning out to be carbon stars instead.  

This paper focuses on Program 200, (P.I.\ Sloan; hereafter
referred to as the MC\_DUST program).  The next section 
describes the sample and the ground-based support
observations.  In \S 3, we describe the IRS observations,
the data reduction, and the classification of the spectra.  
The MC\_DUST sample contains a rich variety of object types, 
each of which requires a different method of analysis.  We 
examine each of these types in turn in the following 
sections:  \S 4---stars with no associated dust (naked stars), 
\S 5---dusty oxygen-rich stars, \S 6---carbon stars, 
\S 7---red sources with emission from polycyclic aromatic 
hydrocarbons (PAHs).  Several targets show characteristics 
which place them in multiple categories, and as a consequence, 
they will be analyzed in more than one of the following 
sections.  In each of these sections, we examine individual 
sources and unusual spectral features, but we save a more 
general comparison of the various populations for \S 8.  
Finally, we draw conclusions in \S 9.

\section{The Sample} 

\subsection{Sample Selection} 

\begin{deluxetable}{lrrrr} 
\tablenum{1}
\tablecolumns{5}
\tablewidth{0pt}
\tablecaption{The MC\_DUST program\label{Tbl2}}
\tablehead{
  \colhead{Object} & \multicolumn{2}{c}{LMC} & \multicolumn{2}{c}{SMC} \\
  \colhead{Class} & \colhead{Planned} & \colhead{Observed} & 
                    \colhead{Planned} & \colhead{Observed} 
}
\startdata
AGB/M         &       7  &       7  &       4  &       4 \\
AGB/MS        &       3  &       3  &       4  &       4 \\
AGB/S         &       3  &       2  &       1  &       1 \\
AGB/C         &       6  &       4  &       4  &       1 \\
OH/IR star    &       3  &       3  & \nodata  & \nodata \\
C/IR star     &       3  &       2  & \nodata  & \nodata \\
RSG           &       3  &       4  & \nodata  &       1 \\
YSO           & \nodata  &       1  & \nodata  & \nodata \\
unknown       & \nodata  &       1  & \nodata  & \nodata \\
total         &      28  &      27  &      13  &      11 \\
\enddata
\end{deluxetable}

\begin{deluxetable*}{lllrrrlrr} 
\tablenum{2}
\tablecolumns{9}
\tablewidth{0pt}
\small
\tablecaption{The MC\_Dust sample\label{Tbl3}}
\tablehead{
  \colhead{ } & \colhead{RA} & \colhead{Dec.} &
  \multicolumn{3}{c}{2MASS} & 
  \colhead{Source} & \colhead{Period} & \colhead{ } \\
  \colhead{Target} & \multicolumn{2}{c}{J2000.0} & \colhead{J} & 
  \colhead{H} & \colhead{K$_s$} & \colhead{Type} & \colhead{(days)} & 
  \colhead{Ref.\tablenotemark{a}}
}
\startdata

HV 11223          & 00 32 01.61 & $-$73 22 34.7 & 11.166 & 10.276 &  9.971 & AGB/MS  &    407: & 1\\
HV 1366           & 00 42 49.87 & $-$72 55 11.4 & 12.232 & 11.431 & 11.154 & AGB/M   &    305  & 5\\
BFM 1             & 00 47 19.24 & $-$72 40 04.5 & 12.661 & 11.647 & 11.027 & AGB/S   &    398  & 5\\
CV 78             & 00 49 03.94 & $-$73 05 19.9 & 11.649 & 10.613 & 10.129 & AGB/C   &    423  & 6\\
HV 11303          & 00 52 08.84 & $-$71 36 24.0 & 10.836 &  9.983 &  9.540 & AGB/MS  &    534  & 1\\
HV 11329          & 00 53 39.40 & $-$72 52 39.2 & 10.911 &  9.987 &  9.647 & AGB/M   &    377  & 5\\
HV 838            & 00 55 38.22 & $-$73 11 41.1 & 10.608 &  9.860 &  9.440 & AGB/M   &    622  & 5\\
HV 11366          & 00 56 54.77 & $-$72 14 08.6 & 11.066 & 10.154 &  9.870 & AGB/MS  &    366  & 1\\
HV 12149          & 00 58 50.17 & $-$72 18 35.6 &  9.963 &  9.022 &  8.609 & AGB/M   &    745  & 5\\
HV 1963           & 01 04 26.63 & $-$72 34 40.3 & 10.638 &  9.677 &  9.409 & AGB/MS  &    249  & 6\tablenotemark{b}\\
Massey SMC 59803  & 01 04 38.21 & $-$72 01 27.0 &  9.093 &  8.301 &  8.097 & RSG     & \nodata & \nodata\\
\\
IRAS 04509$-$6922 & 04 50 40.47 & $-$69 17 31.9 &  9.874 &  8.670 &  7.928 & (AGB/M) &   1290  & 3\\
IRAS 04516$-$6902 & 04 51 29.00 & $-$68 57 50.1 &  9.927 &  8.619 &  7.914 & (AGB/M) &   1170  & 5\\
IRAS 04530$-$6916 & 04 52 45.67 & $-$69 11 49.5 & 13.942 & 11.859 &  9.959 & (RSG)   &   1260: & 3\\
IRAS 04545$-$7000 & 04 54 10.06 & $-$69 55 58.3 & 16.510 & 12.815 & 10.402 & OH/IR   &   1270  & 3\\
HV 888            & 05 04 14.14 & $-$67 16 14.4 &  8.010 &  7.188 &  6.781 & RSG     &    850  & 1\\
HV 2310           & 05 06 27.68 & $-$68 12 03.7 & 10.455 &  9.635 &  9.111 & AGB/M   &    600  & 5\\
IRAS 05128$-$6455 & 05 13 04.57 & $-$64 51 40.3 & 14.553 & 12.832 & 11.281 & (C/IR)  &    708  & 4\\
MACHO 79.5505.26  & 05 14 29.83 & $-$68 54 33.8 & 12.302 & 11.363 & 10.923 & AGB/C   &     97  & 5\tablenotemark{b}\\
HV 5810           & 05 24 07.02 & $-$69 23 36.9 & 10.813 &  9.947 &  9.642 & AGB/M   &    374  & 5\\
WBP 17            & 05 26 19.88 & $-$69 41 37.3 & 12.342 & 11.315 & 10.796 & AGB/C   &    304  & 5\\
WBP 29            & 05 26 40.95 & $-$69 23 11.4 & 12.270 & 11.382 & 10.915 & AGB/C   &    247  & 5\tablenotemark{b}\\
WBP 42            & 05 27 04.73 & $-$69 38 16.3 & 13.800 & 12.192 & 11.009 & AGB/C   &    404  & 5\\
WOH G 339         & 05 27 10.24 & $-$69 36 26.7 & 10.399 &  9.544 &  9.159 & AGB/M   &    538  & 5\\
WBP 77            & 05 28 06.75 & $-$69 32 27.9 & 12.313 & 11.482 & 11.197 & AGB/S   &    212  & 5\\
WBP 104           & 05 28 26.73 & $-$69 14 43.5 & 12.364 & 11.422 & 11.047 & AGB/S   &    245  & 2\\
2MASS J052832     & 05 28 32.10 & $-$69 29 08.5 & 11.429 & 10.571 & 10.326 & unknown &    110: & 5\\
HV 2572           & 05 28 36.71 & $-$69 20 04.1 &  9.795 &  8.836 &  8.447 & AGB/M   &    591: & 6\\
HV 2575           & 05 29 59.98 & $-$67 45 01.4 & 11.482 & 10.556 & 10.066 & AGB/MS  &    390  & 5\\
HV 2578           & 05 29 03.15 & $-$69 48 06.9 &  9.947 &  9.104 &  8.737 & AGB/M   &    653  & 5\tablenotemark{b}\\
IRAS 05300$-$6651 & 05 30 03.89 & $-$66 49 24.2 & 16.895 & 14.800 & 12.247 & C/IR    & \nodata & \nodata\\
HV 996            & 05 32 35.62 & $-$67 55 09.0 &  8.991 &  8.156 &  7.643 & RSG     &    760  & 1\\
IRAS 05329$-$6708 & 05 32 51.33 & $-$67 06 52.0 & 17.285 & 13.638 & 11.233 & OH/IR   &   1260  & 3\\
HV 12620          & 05 32 59.92 & $-$70 41 23.6 & 11.206 & 10.357 & 10.052 & AGB/MS  &    326  & 5\\
IRAS 05348$-$7024 & 05 34 15.99 & $-$70 22 52.5 & 17.659 & 15.488 & 12.851 & C/IR    & \nodata & \nodata\\
IRAS 05402$-$6956 & 05 39 44.89 & $-$69 55 18.1 & 14.344 & 11.740 &  9.789 & OH/IR   &   1390  & 3\\
HV 12667          & 05 49 13.36 & $-$70 42 40.7 & 10.823 & 10.019 &  9.445 & AGB/M   &    645  & 5\\
HV 12070          & 05 52 27.86 & $-$69 14 10.0 & 10.605 &  9.754 &  9.204 & AGB/MS  &    599  & 5\\
\enddata
\tablenotetext{a}{References for periods:  (1) \cite{wbf83}; (2) \cite{wbp85};
  (3) \cite{woo92}; (4) \cite{whi03}; (5) from MACHO data \citep{macho}; 
  (6) from OGLE data \citep{ogle97,ogle05}.}
\tablenotetext{b}{The period has changed significantly from previously 
  published values; see \S 2.3.}
\end{deluxetable*}

The MC\_DUST program was a small (17.2 hour) program 
conducted as part of the IRS Guaranteed Time Observations 
(GTO).  It sampled sources in the LMC and SMC covering a 
range of classifications determined in both the optical and 
the infrared.  The targets were divided into seven classes:  
red supergiants (RSGs), four categories of optically 
identified AGB sources (M giants, MS stars, S stars, and C 
stars), and two categories of infrared sources: OH/IR stars 
and infrared carbon-rich sources (C/IR stars).  Table 1 
breaks down the observing program by object class, as 
planned and as actually executed.  The plan included 28 
sources in the LMC (20 oxygen-rich and eight carbon-rich), 
and 13 in the SMC (nine oxygen-rich and four carbon-rich).  
As with other programs in the Magellanic Clouds, the crowded 
regions resulted in several peak-up failures.  In our case, 
we lost three objects in the LMC and three in the SMC, but 
gained three serendipitious observations, two in the LMC and 
one in the SMC.


The other spectroscopic samples in the Magellanic Clouds 
were selected primarily from near- and mid-infrared 
photometry, but most of the sources in the MC\_DUST sample 
were selected based on optical spectral classifications, 
supplemented with longer wavelength data.  While many of the 
object types in other programs turned out to be different 
than planned, our sources, for the most part, had the 
photospheric chemistries we had expected.  Another advantage 
was that we were able to observe several sources below the 
photometric limits of previous infrared surveys, allowing us 
to probe fainter sources on the AGB.


Table 2 lists the individual targets in the MC\_DUST sample.
The targets with Harvard Variable numbers and the carbon star 
CV 78\footnote{CV = Cordoba Variable; \cite{wbf83} cite 
\cite{des59}.} were selected from the study by \cite{wbf83}.  
These targets account for most of the sources in the RSG, 
AGB/M and AGB/MS classes.  The AGB/S and AGB/C stars with 
WBP numbers were selected from the observations of 
\cite{wbp85}.  \cite{wbf83} identified MACHO 79.5505.26 as 
HV 5680, but HV 5680 is actually 16\arcsec\ to the northeast 
and a magnitude fainter at K.  The target we have named 
BFM 1 was the first S star identified in the SMC \citep{bfm81} 
and, as discussed below, is the only true S star in our 
sample.\footnote{Table 2 corrects the erroneous coordinates
for BFM 1 given by \cite{bfm81}.}  We also 
selected six oxygen-rich infrared sources first studied by 
\cite{woo92}.  They identified three as OH/IR stars.  The 
other three include a suspected supergiant (IRAS 04530) and 
two luminous and embedded evolved stars (IRAS 04509 and 
IRAS 04516).  Finally, we selected three sources believed 
to be infrared carbon stars (C/IR) from a sample studied by 
\cite{vl99}.  The types in parentheses in Table 2 have
changed, as discussed in \S 3.3 and 3.4 below.

Peak-up failures affected our sample of carbon stars most
significantly, reducing our LMC sample from eight to six
carbon stars and our SMC sample from four to one.  
Fortunately, other programs produced a surplus of carbon 
stars, which has resulted in a series of papers on carbon 
stars in the SMC \citep{slo06b,lag07} and the LMC 
\citep{zij06,lei08}.  Three of the peak-up failures resulted 
in serendipitious observations.  Instead of observing NGC 371 
LE 28 in the SMC, we observed Massey SMC 59803, a supergiant 
with a spectral class of K2--3 I \citep{lev06}.  In the LMC, 
we observed the giant WOH G 339 instead of WBP 51, and 
instead of WBP 116, we observed a source identified only in 
the 2MASS catalog (2MASS J05283210-6929084, or J052832 for 
short).  Little is known about this last source; it may even 
be a foreground object and not in the SMC.

\subsection{Ground-based Spectroscopy} 

\begin{figure*} 
\includegraphics[width=6.0in]{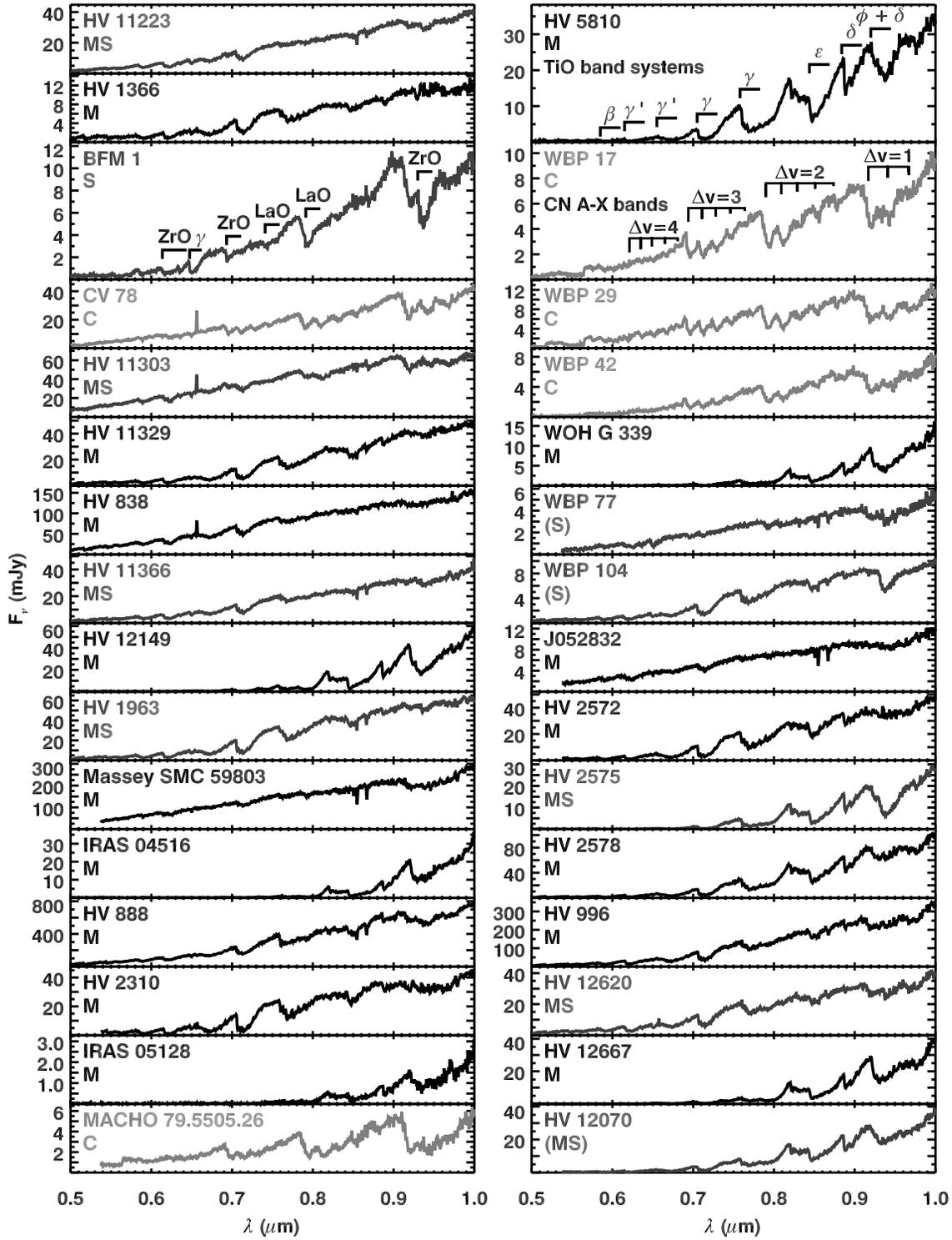}
\caption{Optical spectra of our sample of evolved stars in
the Magellanic Clouds.  The panels for HV 5810, BFM 1, and
WBP 17 are larger than the rest to illustrate typical 
molecular bands in M giants, S stars, and carbon stars, 
respectively.  Spectral classes in parentheses are 
inconsistent with the observed spectra.  We observed only 
two of the nine IRAS sources due to the heavy optical 
obscuration.}
\end{figure*}

To verify the overall chemistry of our targets, we obtained
optical spectra using the Double Beam Spectrograph on the
2.3 m telescope of the Australian National University at 
Siding Spring Observatory (SSO) of all of our targets, except
for seven of the nine IRAS sources, due to their heavy optical 
obscuration.  These spectra covered the 0.45--1.05~\mum\ 
wavelength range with a resolution of 10\AA.  They were 
reduced using standard IRAF procedures with HR 718 as a flux 
standard and the weak-lined giant HD 26169 to remove 
telluric features.  Figure 1 presents the optical spectra.
We have expanded the panels for HV 5810, BFM 1, and WBP 17 to
use them as prototypes for M giants, S stars, and carbon
stars, respectively, and make room for labels for the visible
molecular bands.


Cyanogen (CN) bands dominate the spectra of the carbon stars
and make them easy to distinguish, with its characteristic
corrugations.  The bands which appear in this wavelength 
range arise from the A $^2\Pi$ $-$ X $^2\Sigma^+$ electronic 
transition \citep{hh79}.  The band systems are labelled with 
the change in vibrational state, and in each system the 
vibrational states involved increase to the red.  The 
spectrum of CV 78 includes H$\alpha$ emission at 0.656~\mum.

TiO band systems dominate the spectra of the M giants.  The
identifications of the band systems presented in Figure 1 are 
from \cite{hh79} and the models of M dwarfs by \cite{val98}.  
The spectra in the sample show a wide range of band strengths.
The spectra which are most obscured by dust show only the
reddest bands.  Some of the spectra show H$\alpha$ emission.

Figure 1 shows that the spectrum of one of our planned C/IR 
stars, IRAS 05128, is very similar to IRAS 04516 and clearly 
oxygen-rich.  \cite{tra99} classified it as a carbon star 
based on its \iso/PHOT-S spectrum, but subsequently, 
\cite{mat05} found strong H$_2$O absorption and possible weak 
SiO absorption in its 2.9--4.1~\mum\ spectrum.  Our optical 
spectrum confirms their conclusion that IRAS 05128 is 
oxygen-rich.  \cite{vl08} classify IRAS 05128 
as an M9 giant. 

\cite{bfm81} identified BFM 1 as an S star based on a strong
LaO band at 0.79~\mum.  Our spectrum clearly shows both LaO
and ZrO bands, but these are superimposed on a CN spectrum
(compare BFM 1 to CV 78 just below it).  The strong 
absorption band at 0.93~\mum\ is due to ZrO \citep[][and
references therein]{sw69} and arises from the b' $^3\Pi$ $-$
a $^3\Delta$ electronic transition.  It is close to the head 
of the CN A$-$X $\Delta$v=1 band system, but displaced 
$\sim$0.2~\mum\ to the red, making it easy to identify.
The head of the LaO A $^2\Pi_{3/2}$ $-$ X $^2\Sigma^+$
band system at 0.79~\mum\ \citep{wc78} coincides
with the head of the CN A$-$X $\Delta$v=2 band system, but 
the LaO stands out in two ways.  First, the LaO band makes
the head of the CN band too deep, and second, the A 
$^2\Pi_{1/2}$ $-$ X $^2\Sigma^+$ band system appears at 
0.74~\mum.  Similarly, the ZrO B $^1\Pi$ $-$ X 
$^1\Sigma$ band system strengthens the apparent head of
the CN A$-$X $\Delta$v=3 band system at 0.69~\mum, but 
the ZrO $\gamma$ band system from 0.61 to 0.66~\mum\
makes the identification unambiguous.  The $\gamma$
system arises from the A $^3\Phi$ $-$ X $^3\Delta$ 
transition \citep{pd79}.

BFM 1 is the sole true S star in our sample.  WBP 77 shows
a hint of LaO absorption at 0.79~\mum\ but is otherwise
more consistent with an MS spectrum.  Both it and WBP 104 
show the 0.65 and 0.93~\mum\ ZrO bands seen in many MS 
spectra.  The spectrum of HV 12070, which is classified as an 
MS star, looks more like the spectrum of an M giant.  We have
noted the uncertainties in the classifications by putting
them in parentheses in Figure 1, but we have not changed them
for two reasons.  First, these sources are variable stars
and the spectra may have different characteristics at 
different phases.  Second, we do not distinguish between M
and MS stars in our analysis below; we consider them as one
group of oxygen-rich sources in our analysis.  However, the
fact that BFM 1 is our sole strong S star is significant.
It is the only source where C/O$\approx$1, which may
explain its unusual dust properties (see \S 5.6).

\subsection{Optical Light Curves} 

Many of the stars in our sample have light curves in the 
databases from the MACHO Project \citep[Massive Compact Halo 
Objects;][]{macho} and/or the Optical Gravitational Lensing 
Experiment \citep[OGLE;][]{ogle97,ogle05}.  Table 2 includes 
the periods we have determined from these data.  Most of these 
stars pulsate regularly, or semi-regularly, with periods 
covering a wide range, from $\sim$100 days to $\sim$1400.  
The red and blue lightcurves sometimes give slightly 
different periods.  We have found that the blue light curves 
generally produce more self-consistent periods.

Some of the periods have changed significantly from 
previously published results, primarily because some of the
older periods were based on data which did not sample the
light curves adequately.  Several periods measured by 
\cite{wbf83} have changed.  HV 1963 had a period of 330: 
days, but we measure a period of 249 days with the OGLE data.  
The MACHO data for MACHO 79.5505.26 give a period of 97 days,
compared to 151.  This source has a longer cycle of 1200 
days, so the shorter period must be an overtone mode.  The 
period of HV 2578 has changed from 470 to 247 days, based on 
the MACHO data.  HV 2572 previously had a period of 201 days.  
The OGLE data are poorly characterized, but a period of 591 
days fits the data better than any other.  \cite{wbp85}
measured the period of WBP 29 to be 360 days, but we find that
the MACHO data are more consistent with a 247-day period.
Several other periods have changed by amounts less than 10\%.

\subsection{Near-infrared Photometry} 

\begin{deluxetable*}{lrrrrr} 
\tablenum{3}
\tablecolumns{6}
\tablewidth{0pt}
\small
\tablecaption{Near-infrared SSO photometry\label{Tbl4}}
\tablehead{
  \colhead{ } & \multicolumn{4}{c}{Magnitude} & \colhead{Julian} \\
  \colhead{Target} & \colhead{J} & \colhead{H} & \colhead{K} &
  \colhead{L} & \colhead{Date}
}
\startdata
HV 11223          & 11.197$\pm$0.017 & 10.266$\pm$0.023 & 10.077$\pm$0.013 &  9.673$\pm$0.081 & 2453526\\
HV 1366           & 12.258$\pm$0.019 & 11.534$\pm$0.012 & 11.305$\pm$0.008 & 10.451$\pm$0.196 & 2453334\\
BFM 1             & 12.377$\pm$0.034 & 11.489$\pm$0.018 & 11.082$\pm$0.019 & 10.103$\pm$0.164 & 2453524\\
BFM 1             & 12.088$\pm$0.013 & 11.268$\pm$0.014 & 10.830$\pm$0.014 & 10.186$\pm$0.157 & 2453942\\
CV 78             & 11.099$\pm$0.036 & 10.168$\pm$0.033 &  9.880$\pm$0.012 &  9.223$\pm$0.058 & 2453527\\
HV 11303          & 10.770$\pm$0.022 &  9.846$\pm$0.010 &  9.626$\pm$0.012 &  9.298$\pm$0.067 & 2453530\\
HV 11329          & 11.068$\pm$0.013 & 10.007$\pm$0.013 &  9.800$\pm$0.012 &  9.454$\pm$0.074 & 2453527\\
HV 11329          & 10.991$\pm$0.012 & 10.000$\pm$0.007 &  9.781$\pm$0.018 &  9.752$\pm$0.113 & 2453915\\
HV 838            &  9.824$\pm$0.006 &  9.008$\pm$0.003 &  8.824$\pm$0.004 &  8.607$\pm$0.045 & 2453227\\
HV 11366          & 11.135$\pm$0.016 & 10.115$\pm$0.015 &  9.941$\pm$0.012 &  9.673$\pm$0.085 & 2453530\\
HV 12149          & 10.701$\pm$0.015 &  9.758$\pm$0.009 &  9.353$\pm$0.014 &  8.643$\pm$0.045 & 2453530\\
HV 1963           & 10.606$\pm$0.006 &  9.630$\pm$0.005 &  9.388$\pm$0.003 &  9.233$\pm$0.011 & 2453227\\
HV 1963           & 10.711$\pm$0.011 &  9.735$\pm$0.007 &  9.509$\pm$0.017 &  9.368$\pm$0.079 & 2453915\\
Massey SMC 59803  &  9.056$\pm$0.015 &  8.254$\pm$0.011 &  8.141$\pm$0.012 &  7.952$\pm$0.018 & 2453530\\
\\
IRAS 04509$-$6922 &  9.959$\pm$0.006 &  8.724$\pm$0.011 &  8.098$\pm$0.015 &  7.071$\pm$0.015 & 2453517\\
IRAS 04516$-$6902 & 10.229$\pm$0.008 &  8.943$\pm$0.010 &  8.334$\pm$0.015 &  7.280$\pm$0.017 & 2453517\\
IRAS 04530$-$6916 & 13.869$\pm$0.020 &  \nodata         &  9.768$\pm$0.033 &  7.678$\pm$0.025 & 2453135\\
IRAS 04545$-$7000 & 15.900$\pm$0.320 &  \nodata         & 10.119$\pm$0.006 &  8.083$\pm$0.053 & 2453135\\
IRAS 04545$-$7000 & 14.472$\pm$0.026 & 11.422$\pm$0.007 &  9.397$\pm$0.005 &  7.221$\pm$0.016 & 2453918\\
HV 888            &  7.885$\pm$0.008 &  7.014$\pm$0.009 &  6.788$\pm$0.026 &  6.349$\pm$0.017 & 2453532\\
HV 2310           & 10.756$\pm$0.004 &  \nodata         &  9.533$\pm$0.005 &  8.969$\pm$0.080 & 2453135\\
IRAS 05128$-$6455 & 13.316$\pm$0.010 & 11.451$\pm$0.010 & 10.052$\pm$0.014 &  8.186$\pm$0.032 & 2453531\\
MACHO 79.5505.26  & 12.419$\pm$0.026 &  \nodata         & 10.994$\pm$0.009 &  \nodata         & 2453135\\
HV 5810           & 11.289$\pm$0.009 & 10.307$\pm$0.012 & 10.000$\pm$0.016 &  9.534$\pm$0.135 & 2453530\\
WBP 17            & 12.454$\pm$0.007 & 11.322$\pm$0.011 & 10.792$\pm$0.015 & 10.322$\pm$0.186 & 2453516\\
WBP 29            & 12.555$\pm$0.010 & 11.411$\pm$0.011 & 10.806$\pm$0.020 &  \nodata         & 2453516\\
WBP 42            & 12.797$\pm$0.026 & 11.283$\pm$0.012 & 10.311$\pm$0.015 &  9.168$\pm$0.062 & 2453517\\
WBP 77            & 12.562$\pm$0.031 &  \nodata         & 11.359$\pm$0.004 &  \nodata         & 2453135\\
WBP 104           & 12.387$\pm$0.015 & 11.398$\pm$0.013 & 11.115$\pm$0.015 & 10.641$\pm$0.189 & 2453532\\
HV 2572           & 10.079$\pm$0.008 &  \nodata         &  8.854$\pm$0.006 &  8.418$\pm$0.039 & 2453135\\
HV 2575           & 11.446$\pm$0.007 & 10.433$\pm$0.010 & 10.091$\pm$0.015 &  9.501$\pm$0.089 & 2453530\\
HV 2575           & 11.391$\pm$0.006 & 10.438$\pm$0.004 & 10.117$\pm$0.004 &  9.649$\pm$0.072 & 2453919\\
HV 2578           & 10.085$\pm$0.012 &  9.146$\pm$0.013 &  8.790$\pm$0.015 &  8.199$\pm$0.039 & 2453527\\
IRAS 05300$-$6651 &  \nodata         & 14.972$\pm$0.107 & 12.532$\pm$0.044 &  9.437$\pm$0.086 & 2453526\\
HV 996            &  8.799$\pm$0.007 &  7.910$\pm$0.016 &  7.468$\pm$0.014 &  6.665$\pm$0.014 & 2453526\\
IRAS 05329$-$6708 & 15.096$\pm$0.170 &  \nodata         &  9.794$\pm$0.008 &  7.445$\pm$0.034 & 2453135\\
IRAS 05329$-$6708 & 15.618$\pm$0.074 & 12.557$\pm$0.015 & 10.504$\pm$0.014 &  8.159$\pm$0.023 & 2453915\\
HV 12620          & 11.197$\pm$0.007 & 10.302$\pm$0.011 & 10.104$\pm$0.019 &  9.472$\pm$0.085 & 2453517\\
IRAS 05348$-$7024 & \nodata          &  \nodata         & 13.579$\pm$0.200 &  9.226$\pm$0.080 & 2453135\\
IRAS 05402$-$6956 & 13.632$\pm$0.013 & 11.054$\pm$0.011 &  9.289$\pm$0.015 &  7.170$\pm$0.017 & 2453517\\
HV 12667          & 10.793$\pm$0.006 &  9.853$\pm$0.020 &  9.354$\pm$0.020 &  8.527$\pm$0.037 & 2453521\\
HV 12070          & 10.421$\pm$0.003 &  \nodata         &  9.157$\pm$0.004 &  8.596$\pm$0.041 & 2453135\\
\enddata
\end{deluxetable*}

We obtained ground-based near-infrared photometry for our 
sources as close in time to the IRS observations as possible.
We used CASPIR \citep{caspir} on the 2.3-m telescope at SSO
to obtain images at J (effective wavelength 1.24~\mum), H 
(1.68~\mum), K (2.22~\mum), and narrow-band L (3.59~\mum).
\cite{slo06b} describe the calibration methods.  Table 3
presents the photometry.  It includes two entries
for four sources because we observed them twice with the 
IRS and twice from the ground.  The photometric data for
WBP 17 are averaged from measurements made before and after
the IRS spectroscopy.  


\section{IRS Spectroscopy} 

\subsection{IRS Observations} 

\begin{deluxetable*}{lrrrrlrll} 
\tablenum{4}
\tablecolumns{9}
\tablewidth{0pt}
\small
\tablecaption{Low-resolution observations\label{Tbl5}}
\tablehead{
  \colhead{ } & \colhead{ } & \multicolumn{2}{c}{Time (s)} & 
  \colhead{IRS} & \colhead{Observed} & \colhead{Julian} & \colhead{Infrared} &
  \colhead{ } \\
  \colhead{Target} & \colhead{AOR key} & \colhead{SL} & \colhead{LL} & 
  \colhead{Camp.} & \colhead{(UT)} & \colhead{Date} & \colhead{Sp. Class} &
  \colhead{M$_{bol}$\tablenotemark{a}}
}
\startdata

HV 11223             &  6019584 &  112 &  480 & 21 & 2005 Jun 04 & 2453526 & 1.N       & $-$6.13\\
HV 1366              &  6017024 &  240 &  480 & 14 & 2004 Oct 25 & 2453304 & 1.N       & $-$4.97\\
BFM 1                &  6022912 &  240 &  480 & 21 & 2005 Jun 02 & 2453524 & 2.ST:     & $-$5.03\\
BFM 1 (follow-up)    & 17398784 & 1920 & 2880 & 33 & 2006 Jul 26 & 2453942 & 2.NO      & $-$5.23\\
CV 78                &  6019072 &  112 &  480 & 21 & 2005 Jun 05 & 2453527 & 2.CE      & $-$6.22\\
HV 11303             &  6017536 &  112 &  480 & 21 & 2005 Jun 08 & 2453530 & 1.N:      & $-$6.05\\
HV 11329             &  6017280 &  112 &  480 & 21 & 2005 Jun 05 & 2453527 & 1.NO      & $-$6.25\\
HV 11329 (follow-up) & 17399296 & 1440 & 2880 & 32 & 2006 Jun 28 & 2453915 & 1.NO      & $-$6.27\\
HV 838               &  6017792 &   56 &  480 & 11 & 2004 Aug 11 & 2453228 & 1.N:      & $-$7.37\tablenotemark{b}\\
HV 11366             &  6018304 &  112 &  480 & 21 & 2005 Jun 08 & 2453530 & 1.N:O::   & $-$6.22\\
HV 12149             &  6019328 &   56 &  480 & 21 & 2005 Jun 08 & 2453530 & 2.SE8     & $-$6.69\tablenotemark{c}\\
HV 1963              &  6018048 &  112 &  480 & 10 & 2004 Jul 18 & 2453205 & 1.N:O:    & $-$6.66\\
HV 1963 (follow-up)  & 17399040 &  960 & 1920 & 32 & 2006 Jun 28 & 2453915 & 1.N:O     & $-$6.56\\
Massey SMC 59803     &  6019840 &  240 &  720 & 21 & 2005 Jun 08 & 2453530 & 2.SE4u    & $-$8.11\\
\\
IRAS 04509$-$6922    &  6022400 &   56 &  120 & 21 & 2005 May 26 & 2453517 & 2.SE8f    & $-$7.66\\
IRAS 04516$-$6902    &  6020096 &   56 &  120 & 21 & 2005 May 26 & 2453517 & 2.SE6f    & $-$7.43\\
IRAS 04530$-$6916    &  6023936 &   56 &   56 &  6 & 2004 Apr 18 & 2453114 & 5.U/SA    & $-$7.78\\
IRAS 04545$-$7000    &  6020352 &   56 &  120 &  6 & 2004 Apr 18 & 2453114 & 3.SAxf    & $-$6.42\\
HV 888               &  6015488 &   56 &  120 & 22 & 2005 Jul 10 & 2453562 & 2.SE8tf   & $-$9.01\\
HV 2310              &  6014976 &  112 &  480 &  6 & 2004 Apr 18 & 2453114 & 2.SE5xf   & $-$6.13\\
IRAS 05128$-$6455    &  6024192 &   56 &  240 & 23 & 2005 Aug 09 & 2453592 & 2.SE5f    & $-$5.99\\
MACHO 79.5505.26     &  6016768 &  240 &  480 &  6 & 2004 Apr 16 & 2453112 & 1.N       & $-$4.50\\
HV 5810              &  6014464 &  112 &  480 & 21 & 2005 Jun 08 & 2453530 & 2.U:      & $-$5.64\\
WBP 17               &  6023425 &  240 &  720 & 20 & 2005 Aug 14 & 2453597 & 1.NC      & $-$4.63\\
WBP 29               &  6021632 &  240 &  720 & 20 & 2005 Apr 22 & 2453483 & 2.CE:     & $-$4.54\\
WBP 42               &  6023680 &  112 &  480 & 21 & 2005 May 26 & 2453517 & 2.CE      & $-$4.97\\
WOH G 339            &  6022144 &  112 &  480 & 21 & 2005 May 26 & 2453517 & 2.SE5     & $-$6.52\tablenotemark{d}\\
WBP 77               &  6021120 &  240 &  480 &  6 & 2004 Apr 16 & 2453112 & 2.SE1:    & $-$4.26\\
WBP 104              &  6021376 &  240 &  720 & 22 & 2005 Jul 10 & 2453562 & 1.NO::    & $-$4.53\\
2MASS J052832        &  6020864 &  240 &  720 & 22 & 2005 Jul 10 & 2453562 & 1.NO      & $-$5.45\tablenotemark{d}\\
HV 2572              &  6014208 &   56 &  480 &  6 & 2004 Apr 18 & 2453114 & 2.SE7 tf: & $-$6.78\tablenotemark{e}\\
HV 2575              &  6016256 &  112 &  480 & 21 & 2005 Jun 08 & 2453530 & 1.N:      & $-$5.51\\
HV 2575 (follow-up)  & 17399552 & 1920 & 2880 & 32 & 2006 Jul 02 & 2453919 & 2.NO      & $-$5.51\\
HV 2578              &  6014720 &   56 &  480 & 21 & 2005 Jun 05 & 2453527 & 2.SE7tf   & $-$6.88\\
IRAS 05300$-$6651    &  6024704 &   56 &  240 & 21 & 2005 Jun 04 & 2453525 & 3.CR      & $-$5.17\\
HV 996               &  6015744 &   56 &  120 & 21 & 2005 Jun 04 & 2453526 & 2.SE7     & $-$8.30\\
IRAS 05329$-$6708    &  6023168 &   56 &  120 &  6 & 2004 Apr 18 & 2453114 & 3.SAxf    & $-$7.06\\
HV 12620             &  6016000 &  240 &  720 & 21 & 2005 May 26 & 2453517 & 1.N:      & $-$5.66\\
IRAS 05348$-$7024    &  6024448 &   56 &  480 &  6 & 2004 Apr 16 & 2453112 & 3.CR      & $-$5.62\\
IRAS 05402$-$6956    &  6020608 &   56 &  120 & 21 & 2005 May 26 & 2453517 & 3.SBxf    & $-$7.20\\
HV 12667             &  6015232 &   56 &  480 & 21 & 2005 May 30 & 2453521 & 2.SE6 tf: & $-$6.26\\
HV 12070             &  6016512 &   56 &  120 &  6 & 2004 Apr 16 & 2453112 & 2.SE6     & $-$6.48\\
\enddata
\tablenotetext{a}{Based on {\it Spitzer} spectra and SSO photometry (in Table 3), except where noted.}
\tablenotetext{b}{Using 2MASS and S$^3$MC photometry gives M$_{bol}$ = $-$6.73.}
\tablenotetext{c}{Using 2MASS and S$^3$MC photometry gives M$_{bol}$ = $-$7.38.}
\tablenotetext{d}{Using 2MASS and SAGE photometry to replace absent CASPIR photometry.}
\tablenotetext{e}{Using 2MASS and SAGE photometry gives M$_{bol}$ = $-$7.16.}
\end{deluxetable*}

\begin{deluxetable*}{lrrrlr} 
\tablenum{5}
\tablecolumns{6}
\tablewidth{0pt}
\small
\tablecaption{Short-High observations\label{Tbl6}}
\tablehead{
  \colhead{ } & \colhead{ } & \colhead{SH Int.} & \colhead{IRS} & 
  \colhead{Observed} & \colhead{Julian} \\
  \colhead{Target} & \colhead{AOR key} & \colhead{Time (s)} & \colhead{Camp.} & 
  \colhead{(UT)} & \colhead{Date}
}
\startdata
IRAS 04545$-$7000    & 17399808 & 120 & 32 & 2006 Jul 01 & 2453918.4 \\
IRAS 05329$-$6708    & 17400320 &  60 & 32 & 2006 Jun 28 & 2453914.9 \\
\enddata
\end{deluxetable*}

We observed all of the sources in our sample with the 
low-resolution modules on the IRS.  These observations were 
carried out as part of the IRS GTO allocation, which was 
spread over the first 2.5 years of the {\it Spitzer}
mission.  As described below, we also followed 
up some targets in Cycle 3.  We observed four stars with
possible low-contrast excesses with longer integration 
times (PID 30332), and we followed up two sources in the 
MC\_DUST program with Short-High (SH) observations to verify 
the existence of new dust features at 14~\mum\ (PID 30345).
Table 4 summarizes the low-resolution observations, and 
Table 5 summarizes the SH observations.


\subsection{Standard Reduction} 

Our data reduction began with the S15 version of the data 
pipeline released by the {\it Spitzer} Science Center (SSC).
Before extracting spectra from the flatfielded images, we
subtracted background images and corrected bad pixels.
For the Short-Low (SL) data, we used images with the source
in the other aperture as the background (aperture
differences).  Thus, the Short-Low order 1 (SL1) images 
served as background data for the SL order 2 (SL2) images, 
and vice versa.  For Long-Low, we obtained nod differences,
subtracting images with the source in one nod position from
images with the source in the same aperture, but in the
other nod position.  The next step was to correct bad
pixels by interpolating from their neighbors, using the
{\rm imclean} IDL package\footnote{Available from the SSC
as {\rm irsclean}.}.  Bad pixels are those flagged in the
bit-mask images provided with the data release and those
flagged as rogues in the rogue masks provided by the SSC.

We extracted spectra from the corrected images using the
{\rm profile}, {\rm ridge}, and {\rm extract} routines
available in the SSC's {\it Spitzer} IRS Custom Extractor
(SPICE).  The spectra were calibrated with spectral
corrections generated from observations of the K0 giant 
HR 6348 for SL and for LL, HR 6348 along with the late
K giants HD 166780 and HD 173511 \citep[see][for further
details]{slo08c}.  To eliminate pointing-induced
discontinuities between the orders, we normalized the
various segments to match at the wavelengths where they
overlapped.  We normalized segments upwards to the
presumably best-centered segment.  Finally, we trimmed
the segments to eliminate untrustworthy data at the ends.

The SH observations required a different reduction method.
We used S14 images, substracted dedicated sky images, and
performed full-slit extractions.  We calibrated using
observations of $\xi$ Dra (K2 III).

\subsection{Bolometric Magnitudes} 

Table 4 also includes the bolometric magnitudes of our
sources at the time of observation, which result from an 
integration of the IRS spectra and the near-infrared SSO 
photometry.  We have assumed a Wien tail to the blue and a 
Rayleigh-Jeans tail to the red, and a distance modulus of 
18.5 to the LMC and 18.9 to the SMC.  

Where possible, we have also computed bolometric magnitudes 
by replacing the SSO data with the 2MASS photometry in Table 2 
and 3.6 and 4.5~\mum\ {\it Spitzer} photometry from
the SAGE survey of the LMC \citep{mei06} and the S$^3$MC
survey of the SMC \citep{bol07}.\footnote{SAGE and S$^3$MC
stand, respectively, for Surveying the Agents of Galactic
Evolution and the {\it Spitzer} Survey of the SMC.}  In 
those cases where the data give a significantly different 
result, Table 4 includes a note of explanation.

The observational distinction between red supergiants and 
oxygen-rich AGB stars is non-trivial.  Sources in the 
Magellanic Clouds have known distances, making luminosity
a logical discriminant (pulsation amplitude is another).  
\cite{wbf83} estimate that the classic upper limit for AGB 
sources lies at about M$_{bol}$=$-7.1$, but as \cite{woo92} 
discussed, some AGB sources can occasionally stray above that 
limit.  The bolometric luminosities in Table 4 suffer from 
the additional difficulty of variability.  We have observed 
the stars at only one phase, which blurs the $-7.1$-magnitude 
limit somewhat.

For the classifications in Tables 1 and 2, we assume all RSGs 
exceed the classic limit and show no large-amplitude 
pulsation, OH maser emission, or Li absorption at 0.67~\mum.  
The Li absorption indicates that the source has experienced 
hot-bottom burning, placing it at the high-mass end of the 
AGB \citep{smi95}.  

Four of our targets meet all of these constraints.  HV 888 
and HV 996 are firmly identified supergiants in the LMC 
\citep[e.g.][]{wbf83}, as is Massey SMC 59803 in the SMC 
\citep{lev06}.  The fourth source is IRAS 04530, but this
source is probably a young stellar object (YSO), as 
explained in \S 7.3.

Two more sources meet some, but not all, of the criteria for 
a supergiant.  IRAS 04509 and IRAS 04516 have bolometric 
luminosities over the AGB limit by our measurements (with 
both SSO and 2MASS photometry), but \cite{woo92} suggested 
that they were AGB sources, having measured their bolometric 
luminosities to be $-7.30$ and $-6.77$, and
their amplitudes at K to be 1.5 and 1.3 magnitudes, 
respectively.  They noted that very red AGB sources might 
exceed the $-7.1$-magnitude limit slightly.  However, they 
found no evidence of OH maser activity in IRAS 04509, and 
\cite{mar04} found none in IRAS 04516.  Our bolometric 
luminosities are well above the classic AGB limit, and
while both sources have no detected maser activity, their
K-band amplitudes are inconsistent with supergiant status.
These two sources are as luminous as AGB sources can be, and
we suspect that they are massive AGB sources.

\cite{woo92} and \cite{whi03} detected masers in all three
targets identified as OH/IR sources in Table 1.  IRAS 05329 
and IRAS 05402 have bolometric luminosities in some 
references over the AGB limit, but the presence of maser 
emission clarifies their status.  IRAS 04545 is under the AGB 
limit in most references.

Two sources show evidence of hot-bottom burning, HV 2572
\citep{smi95,wf00} and HV 12070 \citep{whi03}, making them
massive AGB stars.  \cite{wbf83} suggested that HV 2572
might be a foreground object, but its radial velocity is
more consistent with membership in the LMC (P.\ Wood,
private communication).



\subsection{Infrared Spectral Classification} 

The first step in our analysis is to classify each source 
into one of five broad categories based on the general 
properties of its infrared spectrum.  Each spectrum is then 
analyzed in more detail, using a method appropriate to its
classification.

We applied the classification method developed at 
Hanscom Field and introduced by \cite{kra02}.  This system 
places a spectrum into one of five numbered groups based on 
its overall color:  1 for blue spectra dominated by a stellar
continuum, 2 for dusty stars, 3 for spectra dominated by warm 
dust, and 4 and 5 for spectra dominated by progressively 
cooler dust.\footnote{These last two groups are distinguished
using data from 36 to 45~\mum, which are unavailable to the
IRS.  We simply classify the one red spectrum as group 5.}  
It then describes the dominant features in the spectra with 
one- or two-letter abbreviations:  N for naked stars, SE for 
silicate emission, SB for self-absorbed silicate emission, SA 
for silicate absorption, CE for carbon-rich dust emission, CR 
for optically thick and cooler carbon dust shells, and U for 
spectra showing the unidentified infrared (UIR) emission 
features arising from polycyclic aromatic hydrocarbons (PAHs) 
and related material.  We introduce a new classification, 
2.ST, in \S 5.6.

Table 4 gives the infrared spectral classifications, which
include the basic classification plus additional subdivisions
and notes as described in the following sections.  The 
overall classification determines the method(s) applied to 
analyze the spectra and in which sections they are
discussed.  In \S 4, we discuss the naked stars (1.N and 
1.NO) and some of the nearly naked spectra showing other dust 
features.  In \S 5 we examine oxygen-rich dust spectra (2.SE, 
3.SE, 3.SB, and 3.SA), while \S 6 examines carbon-rich 
spectra (1.NC, 2.CE and 3.CR).  Finally, \S 7 looks at the 
spectra with red continua and/or contributions from PAHS.

Distinguishing the nature of the dust (oxygen-rich vs.
carbon-rich) is straightforward if there is plenty of dust,
but for sources with low-contrast excesses, not only can it
be difficult to determine what kind of dust is present, but
it can be a challenge just to decide if there is an excess
at all.  To assist, we have measured the dust emission 
contrast (DEC), which \cite{sp95} defined to be the 
ratio of dust emission to stellar emission between 7.67 and 
14.03~\mum.  They applied this method to the database of
Low-Resolution Spectra (LRS) from {\it IRAS}.  \cite{sp98}
noted that all sources with DEC $>$ 0.08 had clear dust
excesses, while all sources with DEC$<$ 0.04 looked naked.  
The sources between these two limits were more problematic  
and could be described as ``fig-leaf'' sources, since they 
are neither naked nor fully clothed by a dust shell.

We follow their example, but with some changes.  As done 
previously, we assume the optically thin case, which implies 
that the contributions to the spectrum from the star and dust 
are additive.  The better coverage to shorter wavelengths of 
the IRS compared to the LRS allows us to shift the wavelength 
range for fitting the stellar contribution to 6.8--7.4~\mum, 
between an absorption band from water vapor at 
$\sim$6.6~\mum\ and the fundamental SiO band, which begins at 
7.4~\mum.  As discussed below (\S 4), the SiO band is much 
weaker in the Magellanic Clouds than in the Galaxy, so 
instead of using a model spectrum with an SiO band, we simply 
fit and subtract a 3600 K Planck function to measure the DEC.
This also makes the method general enough to apply to carbon 
stars.

While \cite{sp98} set the minimum DEC for dusty stars to be 
0.08, we find that in the MC\_DUST sample, sources with a DEC
as high as 0.10 do not have a sufficient excess to identify
unambiguously.  We classify sources with DEC$\le$0.04 as 1.N,
while the fig-leaf sources with 0.04$<$DEC$\le$0.10 are 1.N:.
Sources with higher dust emission contrasts are classified in 
group 2, and in all cases but one, can be placed into the 
appropriate sub-category (2.SE, 2.CE, etc.).

Three sources (BFM 1, HV 5810, and WBP 77) have DECs 
high enough to qualify as dusty, but we can see through the
dust and investigate their photospheres.
Consequently, we will examine their photospheric properties
in the section below on naked stars (\S 4) and their dust 
content in the appropriate section as well.  

Two of our sources are identified as C/IR stars in Table 2,
IRAS 05300 and IRAS 05348.  \cite{vl06} list
IRAS 05348 as ``unknown type'', but our IRS data clearly show 
a carbon star.  As noted in \S 2.2, IRAS 05128 has previously 
been identified as carbon-rich based on its {\it ISO} 
spectrum, but both our optical and infrared spectra reveal an 
oxygen-rich source.

One source, MACHO 79.5505.26, has inconsistent optical and
infrared spectra.  The optical spectrum is clearly 
carbon-rich, but the infrared spectrum shows no dust or
absorption bands indicating its chemistry, which is unusual
for carbon stars.  For this reason, we analyze it both in
\S 4 and 6.  Similarly, the SMC supergiant Massey
SMC 59803 shows both silicate and PAH emission (analyzed 
separately in \S 5 and 7).

\section{Naked Stars} 

\begin{figure} 
\includegraphics[width=3.5in]{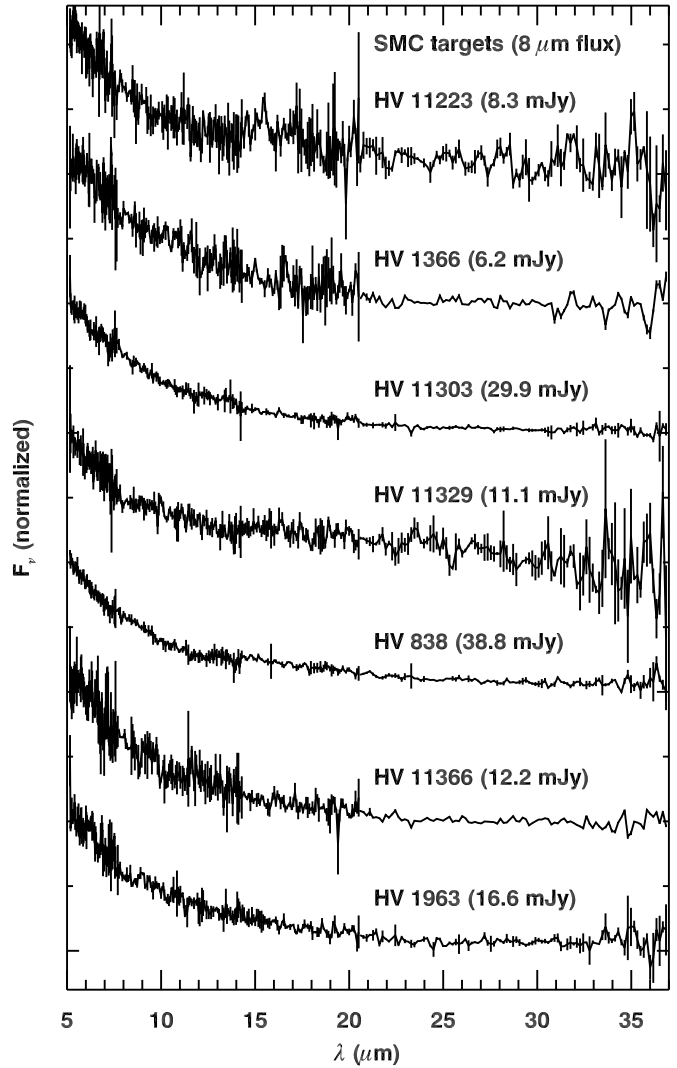}
\caption{IRS spectra of naked oxygen-rich stars in the SMC.  
The spectra have been normalized and shifted for clarity.  
The 8~\mum\ flux, determined by applying the bandpass of the
IRAC 8~\mum\ filter to the IRS data, is given after each 
source name in parentheses.}
\end{figure}

\begin{figure} 
\includegraphics[width=3.5in]{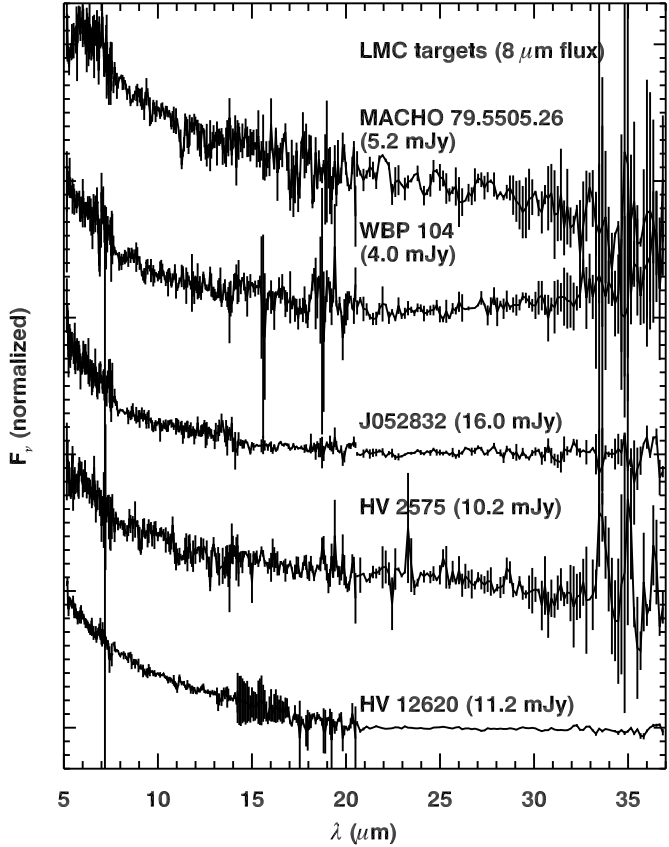}
\caption{IRS spectra of naked oxygen-rich stars in the LMC.  
As in Fig.\ 2, the spectra have been normalized and shifted.}
\end{figure}

Figures 2 and 3 present the spectra of the naked
sources in the SMC and LMC, respectively.  These include seven 
of the 11 SMC sources and five of the 27 LMC sources.  


Many of these spectra are faint, which exposes us to
background issues that would have little effect on brighter
sources.  The Magellanic Clouds are crowded in the infrared,
with extended emission and a multitude of stray sources.
These have a greater impact on LL, with its 10\arcsec\ slit
and sensitivity to extended emission from cool dust.  In some 
of the spectra, the source is faint enough and blue enough 
that we failed to detect it in LL.  For this reason, our 
analysis will concentrate on the SL data.  Even though some 
of the LL data are excellent, we did not detect any 
significant spectral features beyond about 15~\mum.

\subsection{Method} 

\begin{figure} 
\includegraphics[width=3.5in]{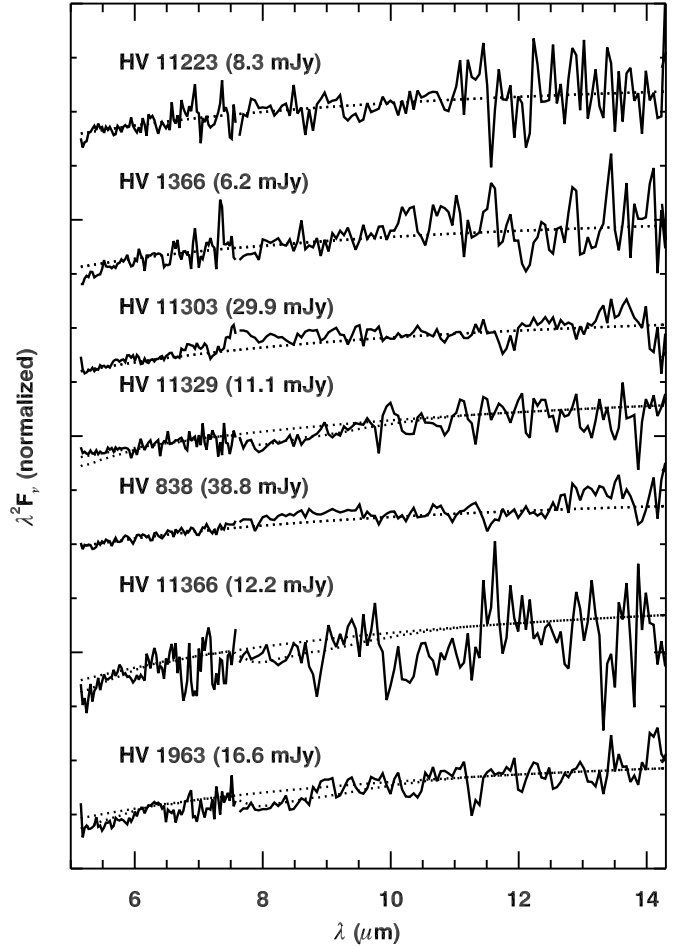}
\caption{IRS spectra of naked stars in the SMC, plotted in 
Rayleigh-Jeans units ($\lambda^2 F_{\nu}$), so that the 
Rayleigh-Jeans tail of a blackbody would be a horizontal 
line.  For each spectrum, the dotted lines give the fitted 
continuum and the best fit absorption from an average 
Galactic M giant.  Table 6 gives the equivalent width of
the fitted SiO band.}
\end{figure}

\begin{figure} 
\includegraphics[width=3.5in]{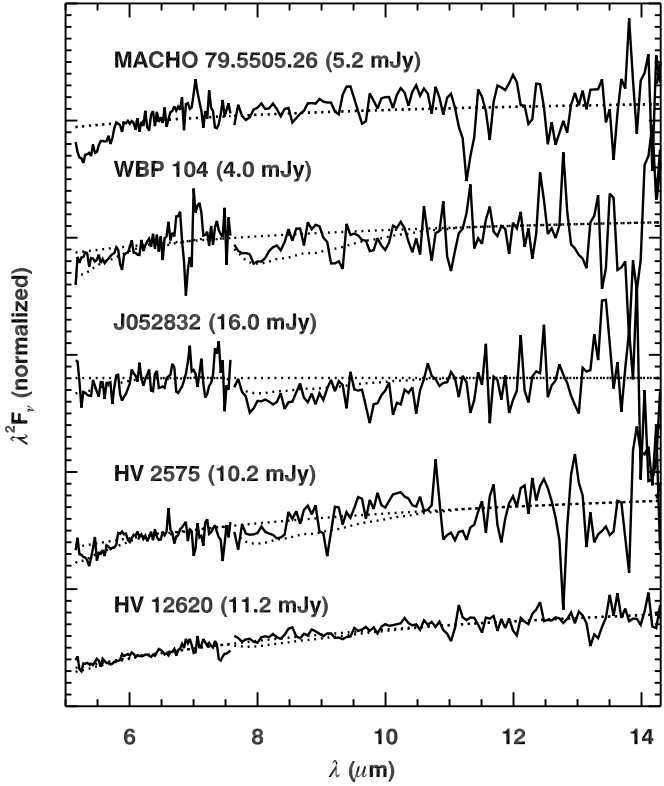}
\caption{IRS spectra of naked stars in the LMC, plotted in 
Rayleigh-Jeans units with the fitted fitted continuum and 
absorption spectra, as in Fig.\ 4.}
\end{figure}

\begin{figure} 
\includegraphics[width=3.5in]{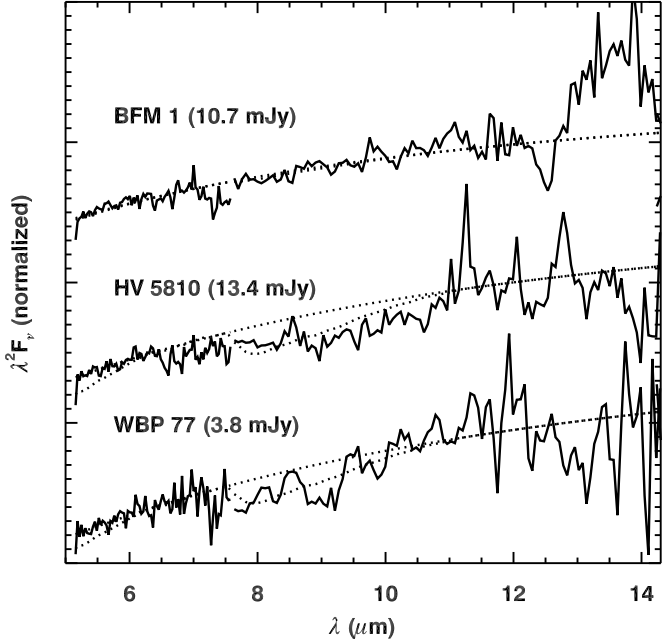}
\caption{IRS spectra of stars in the SMC and LMC with
indications of low-contrast dust emission (in Rayleigh-Jeans
units).  The dotted lines are the fitted continua and 
absorption spectra, as in Fig.\ 4.  Spectra with full
wavelength coverage for these sources appear in later
figures:  BFM 1 in Fig.\ 8, HV 5810 in Fig. 22, and WBP 77
in Fig.\ 9.}
\end{figure}

\begin{deluxetable*}{lrrrl} 
\tablenum{6}
\tablecolumns{5}
\tablewidth{0pt}
\tablecaption{Naked and nearly naked stars\label{Tbl7}}
\tablehead{
  \colhead{ } & \colhead{Dust Em. } & \colhead{Fitted} &
  \colhead{W$_{\lambda}$ SiO} & \colhead{Infrared} \\
  \colhead{Target} & \colhead{Contrast} & \colhead{T (K)} &
  \colhead{(\mum)} & \colhead{Sp. Class}
}
\startdata
HV 11223             & $-$0.02 $\pm$ 0.08 &  2290 $\pm$  510    & $<$ 0.18        & 1.N \\
HV 1366              &    0.02 $\pm$ 0.06 &  2270 $\pm$  560    & $<$ 0.20        & 1.N \\
BFM 1                &    0.24 $\pm$ 0.04 &  1460 $\pm$  110    & $<$ 0.16        & 2.ST: \\
BFM 1 (follow-up)    &    0.16 $\pm$ 0.02 &  1160 $\pm$   10    & 0.15 $\pm$ 0.05 & 2.N:O \\
HV 11303             &    0.07 $\pm$ 0.03 &  2130 $\pm$  240    & $<$ 0.03        & 1.N: \\
HV 11329             &    0.04 $\pm$ 0.05 &  1600 $\pm$   80    & 0.28 $\pm$ 0.13 & 1.NO \\
HV 11329 (follow-up) &    0.00 $\pm$ 0.02 &  2380 $\pm$  100    & 0.18 $\pm$ 0.03 & 1.NO \\
HV 838               &    0.10 $\pm$ 0.03 &  2320 $\pm$  320    & $<$ 0.02        & 1.N: \\
HV 11366             &    0.06 $\pm$ 0.09 &  2230 $\pm$  300    & 0.18 $\pm$ 0.19 & 1.N:O:: \\
HV 1963              &    0.07 $\pm$ 0.05 &  2160 $\pm$  120    & 0.21 $\pm$ 0.13 & 1.N:O: \\
HV 1963 (follow-up)  &    0.05 $\pm$ 0.01 &  1990 $\pm$   50    & 0.13 $\pm$ 0.03 & 1.N:O \\
\\
MACHO 79.5505.26     & $-$0.06 $\pm$ 0.04 &  4120 $\pm$ 2170    & $<$ 0.11        & 1.N \\
HV 5810              &    0.19 $\pm$ 0.08 &  1210 $\pm$  110    & 0.38 $\pm$ 0.17 & 2.U \\
WBP 77               &    0.19 $\pm$ 0.07 &  1130 $\pm$  120    & 0.33 $\pm$ 0.18 & 2.SE1: \\
WBP 104              & $-$0.11 $\pm$ 0.06 &  4080 $\pm$  890    & 0.20 $\pm$ 0.19 & 1.NO:: \\
2MASS J052832        & $-$0.21 $\pm$ 0.05 & 10000 $\pm$ \nodata & 0.34 $\pm$ 0.15 & 1.NO \\
HV 2575              &    0.07 $\pm$ 0.07 &  2170 $\pm$  870    & 0.09 $\pm$ 0.17 & 1.N: \\
HV 2575 (follow-up)  &    0.07 $\pm$ 0.02 &  2260 $\pm$  180    & 0.24 $\pm$ 0.08 & 1.N:O \\
HV 12620             &    0.06 $\pm$ 0.03 &  1900 $\pm$   10    & $<$ 0.04        & 1.N: \\
\enddata
\end{deluxetable*}

The Hanscom system includes two subclasses of naked stars, 
oxygen-rich (1.NO), and carbon-rich (1.NC).  The 1.NO spectra 
can be distinguished by the presence of the SiO fundamental 
band, with maximum absorption at 8.0~\mum.  For 1.NC spectra, 
the primary discriminant is the acetylene (C$_2$H$_2$) 
absorption band at 7.5~\mum, and in some cases, a narrower 
acetylene band at 13.7~\mum.  Those sources which cannot be
identified as 1.NO or 1.NC are classified simply as 1.N.
We examine the one 1.NC spectrum, of WBP 17, in \S 6.

The SiO fundamental has a bandhead at 7.4~\mum.  While the
wavelength of maximum absorption is nearby, at 8.0~\mum, the
band extends out as far as $\sim$11.0~\mum\ in some cases.  
The band is broad enough, and the spectra are noisy enough, 
that our usual technique of fitting a line segment to the
continuum on either side of the band to integrate the
equivalent width is inadequate.  First, the continuum is not 
linear over such a broad wavelength range, and second, our
estimated equivalent width has proven too sensitive to the
vagueries of the continuum fit.  Instead, we have constructed
a template transmission spectrum from Cycle 4 IRS GTO 
observations of 10 naked M giants \citep[PID 40112;][]{slo08d}
and fitted it to the spectrum to estimate the equivalent
width.\footnote{The sample currently includes 13 reduced 
spectra, but we did not use three with strong water vapor 
absorption at 6.6~\mum\ when generating the SiO template.}   
We estimate the continuum with a Planck function which 
reproduces the color measured over the wavelength intervals 
6.10--6.45 and 11.5--13.0~\mum.  We then scale the 
transmission spectrum until we find the minimum $\chi^2$ 
difference, and measure the equivalent width of our fitted 
spectrum from 7.4 to 11.5~\mum.  We estimate the uncertainty 
in the fitted temperature by varying the long-wavelength 
fitting interval, and we estimate the uncertainty in 
equivalent width by finding the SiO strengths which 
correspond to the minimum $\chi^2$ multiplied by the factor 
$(1 + N^{-1/2})$, where $N$ is the number of pixels in the 
measured wavelength range.  

Figures 4 and 5 illustrate the best fits to the naked stars, 
and Figure 6 shows the three sources with both possible 
low-contrast dust and photospheric absorption bands.  The 
results of these fits appear in Table 6.


\subsection{Results} 

\begin{figure} 
\includegraphics[width=3.5in]{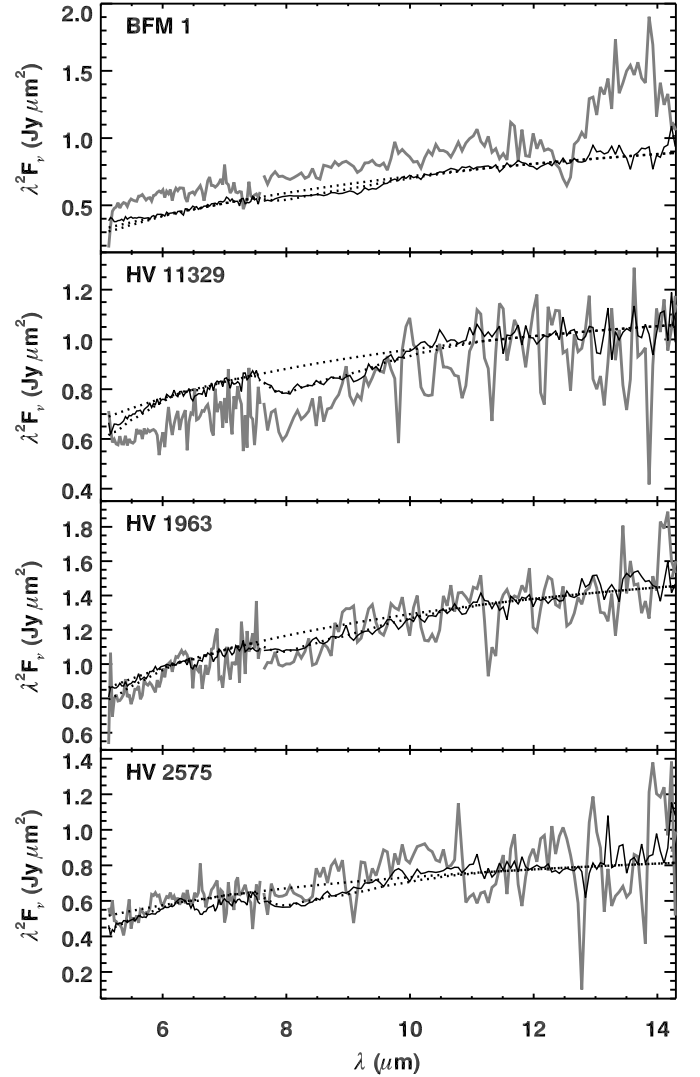}
\caption{The follow-up spectra with longer integration times 
of four of the MC\_DUST targets.  The originally observed
spectrum appears in light gray, while the new spectrum is in 
black.  The dotted lines are the continuum fitted to the new 
spectrum and the best-fit absorption spectrum, as in Fig.\ 3.}
\end{figure}

For each naked spectrum, Table 6 lists the dust emission 
contrast, the temperature of the fitted Planck function, and 
the equivalent width of the best-fitting SiO band.  One 
should not overinterpret the temperatures.
Spectra from sources which were not properly centered in the 
slit will be subtly distorted, since the slit throughput is 
a complex function of both wavelength and position in the 
slit \citep{ks06}.  As Table 6 shows, the uncertainty in the
temperature can be substantial in the noisier spectra.
However, the temperatures can be useful if treated carefully.
Table 6 reveals a correlation between higher dust emission 
contrasts and lower temperatures, suggesting that sources 
with lower blackbody temperatures do show low-contrast excess 
emission from dust.

The classifications of naked stars in Tables 3 and 6 require
some clarification.  If a colon follows the ``N", it indicates
uncertainty about whether or not the source is naked (i.e.\ it
is a ``fig-leaf'' source).  If the colon follows the ``O'',
then the SiO band is detected with a S/N between $\sim$1 and 
$\sim$2.

Figure 6 presents the three sources with a DEC too high to
be naked and too low to identify the nature of the dust
excess.  BFM 1, the S star in the SMC, has a DEC of 
$\sim$0.20, indicating the presence of some dust.  The full
spectrum of this source appears in \S 5, along with a 
discussion of the interesting feature at 13--14~\mum.  
HV 5810 shows an apparent SiO absorption feature, but it 
also shows apparent emission features at 11.3, 12.0, and 
12.7~\mum, the positions of the out-of-plane C--H bending 
modes in PAHs \citep[e.g.][]{atb89}.  The apparent SiO 
absorption is more likely just the trough between the 
7--9~\mum\ and 11--13~\mum\ PAH complexes (despite a S/N
ratio of $\sim$2).  This interpretation is strengthened by 
the spectral structure at 8.5~\mum, which resembles the 
emission feature from the C--H in-plane bending mode in PAHs.
We discuss this source further in \S 7.2 below.


Figure 7 presents the follow-up observations with longer
integrations of four of our targets.  We have detected SiO 
bands in all four of the follow-up spectra at a S/N of 
$\sim$3 or better, which is an improvement over the original 
spectra, where the limited S/N allowed only two detections
with a S/N ratio of just $\sim$2.  The follow-up spectrum 
of BFM 1 has changed compared to the original.  The apparent 
13--14~\mum\ emission feature is absent, while the SiO band
is now clearly present, though weak.  The S/N ratio of the
spectrum of HV 2575 is high enough to reveal weak absorption 
between 6.5 and 7.4~\mum\ from water vapor \citep{cam02}.


We will return to the naked stars in \S 8.1 and 8.3 below.

\section{Oxygen-rich Dust Sources} 

\begin{figure} 
\includegraphics[width=3.5in]{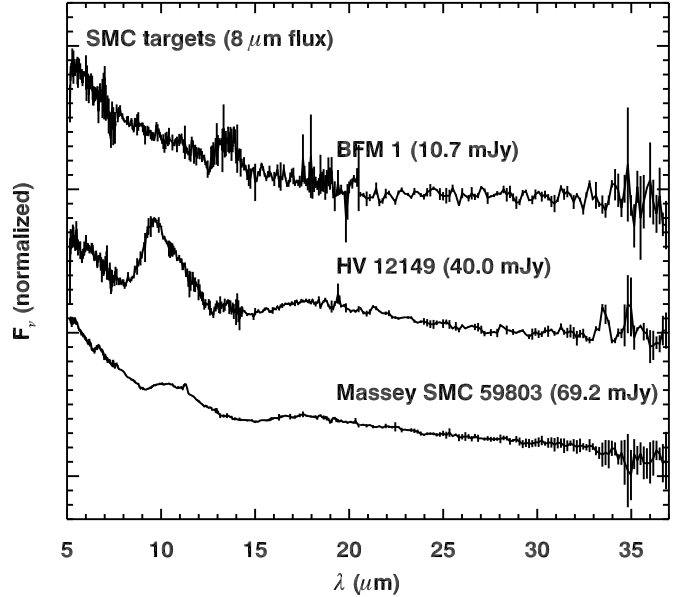}
\caption{Spectra of the three SMC sources showing 
silicate or possibly related dust emission features.}
\end{figure}

\begin{figure*} 
\includegraphics[width=7.0in]{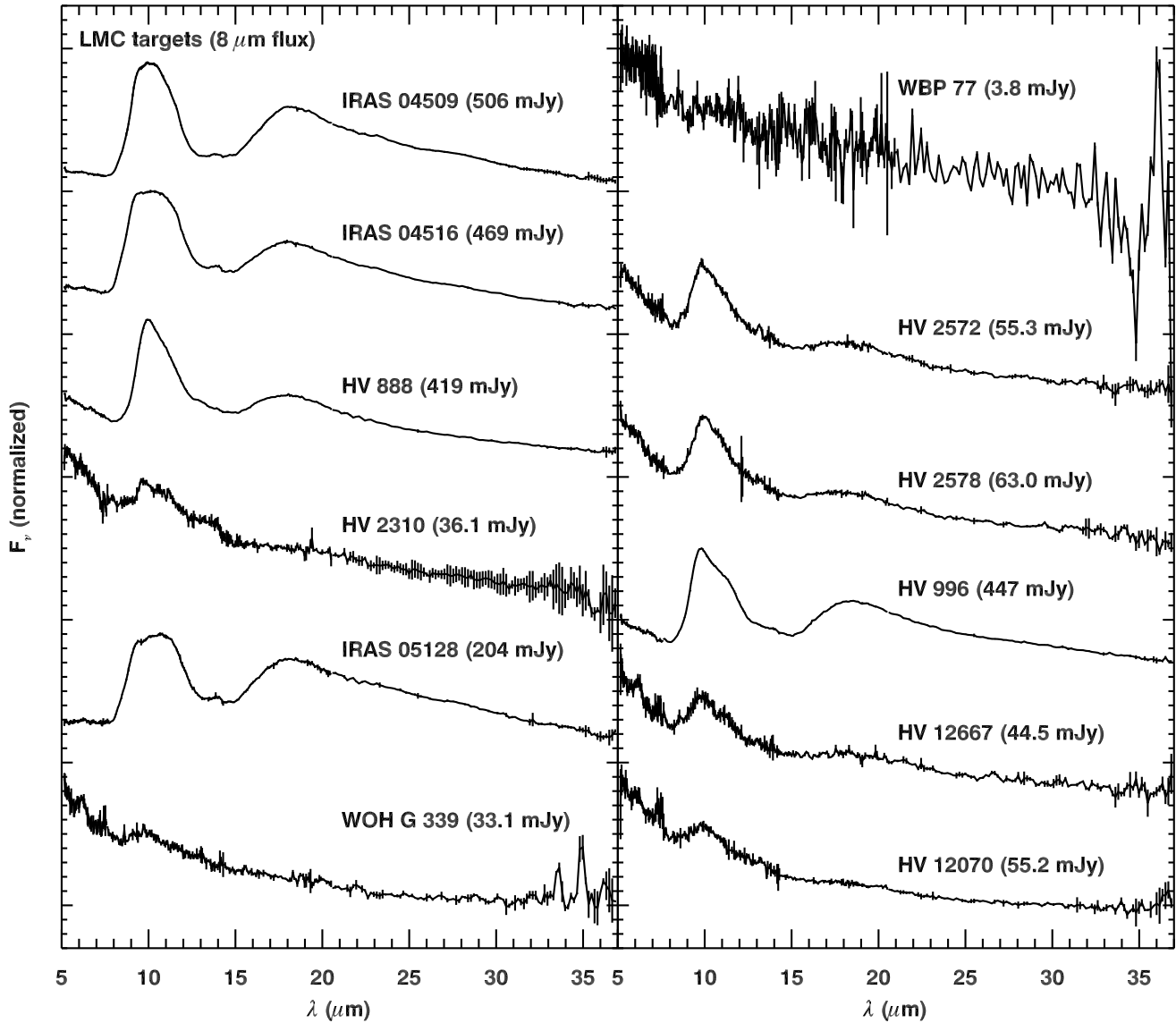}
\caption{Spectra of the 12 LMC sources with silicate or
related emission.}
\end{figure*}

\begin{figure} 
\includegraphics[width=3.5in]{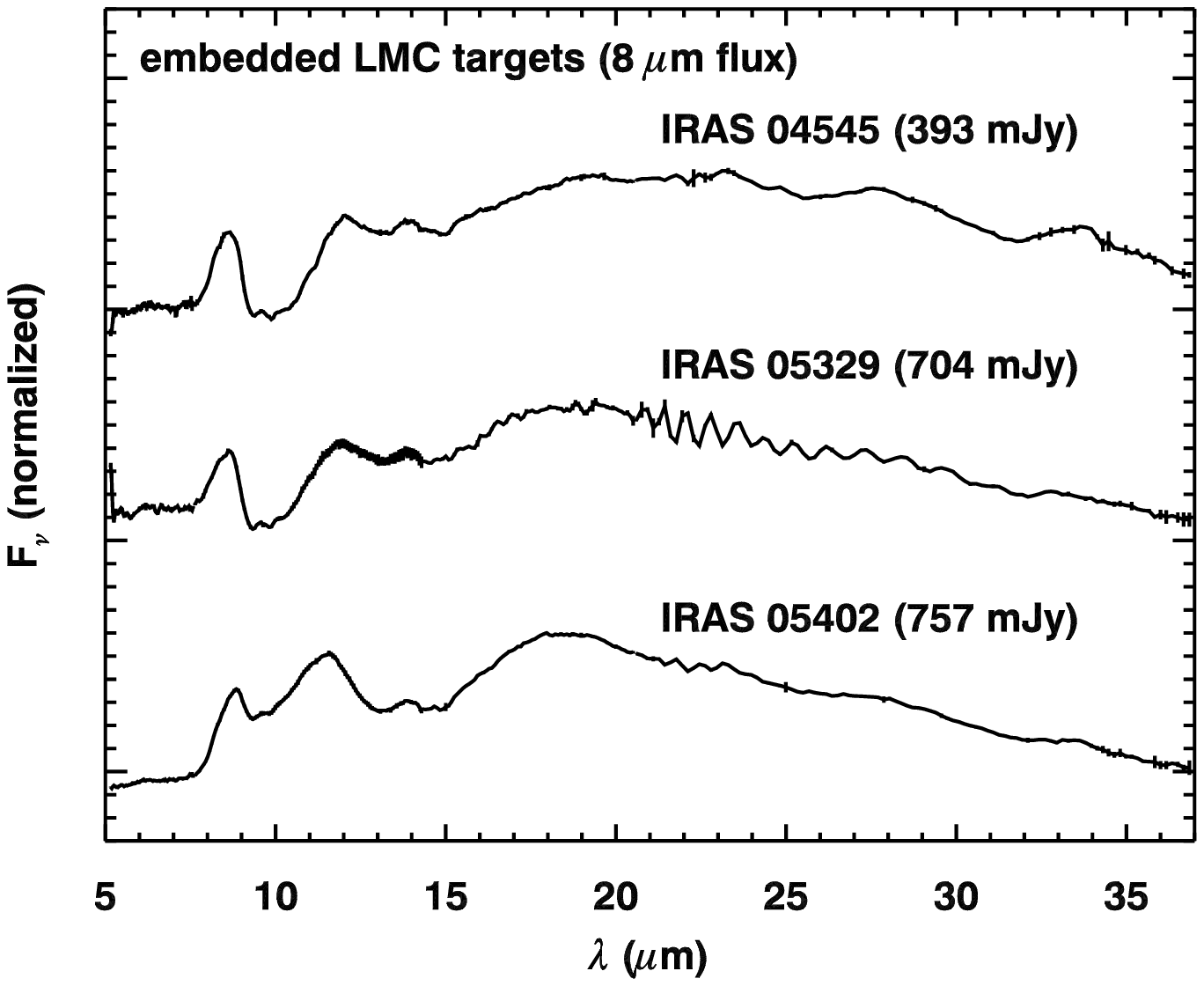}
\caption{Spectra of the three LMC sources showing silicate
absorption features or self-absorbed silicate features at
10~\mum.  These spectra are analyzed in Fig.\ 12 and 13.}
\end{figure}

The MC\_DUST sample contains 18 spectra dominated by silicate 
dust and related oxygen-rich dust species, as Figures 8 through 
10 show.  The SMC sample has three oxygen-rich dust emission 
sources, and no sources with enough dust opacity to drive the 
10~\mum\ feature into absorption.  The LMC sample includes 12 
emission sources, one red source showing a self-absorbed 
10~\mum\ silicate feature, and two even redder sources with 
silicate absorption at 10~\mum.  


\subsection{Classification} 

\begin{figure} 
\includegraphics[width=3.5in]{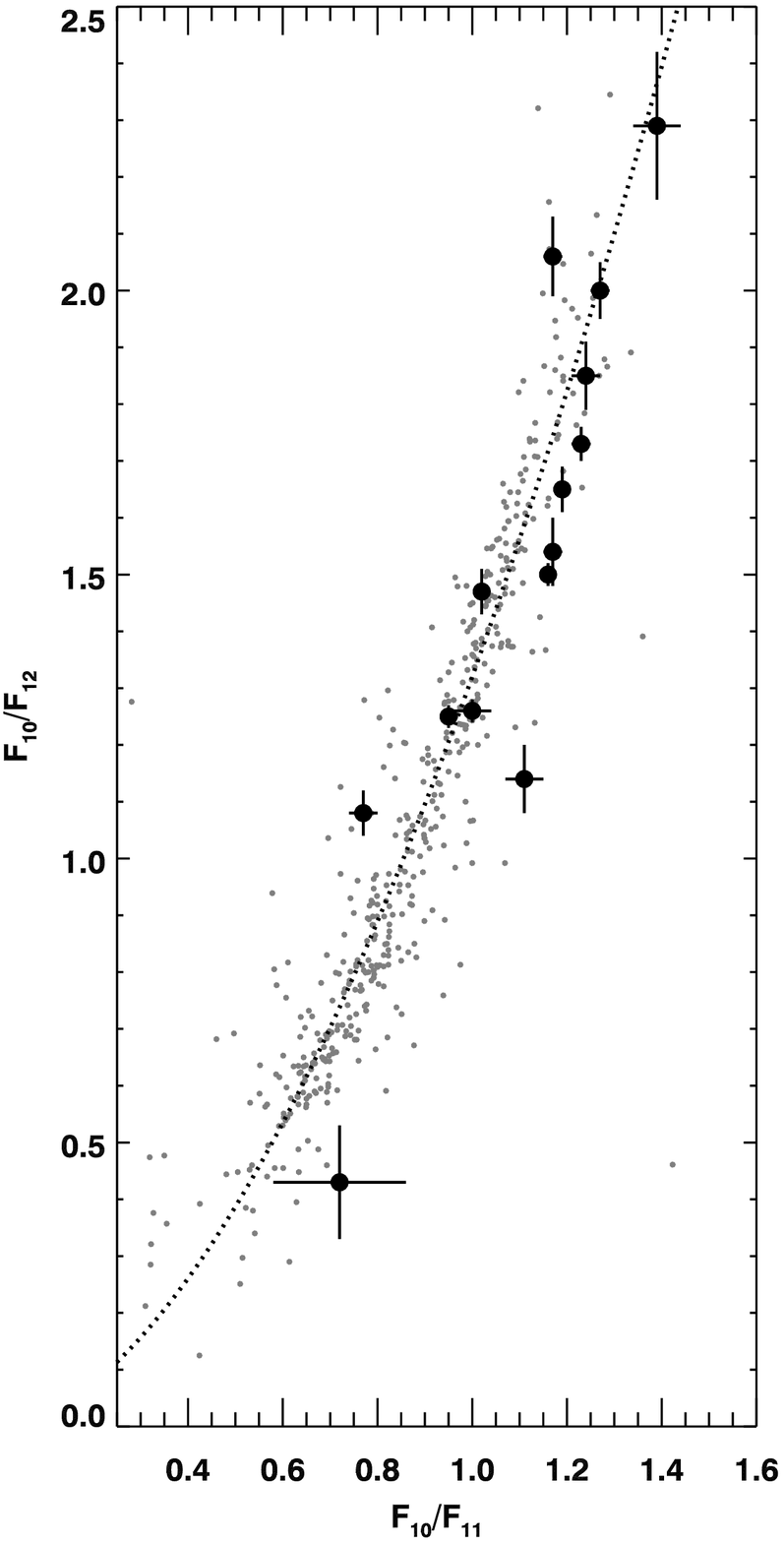}
\caption{The silicate dust sequence.  MC\_DUST sources are
plotted with black filled circles.  The smallar gray circles
show the distribution of Galactic AGB sources 
\citep{sp95,sp98}, and the dotted curve is a power law:  
  $F_{10}/F_{12} = 1.32 (F_{10}/F_{11})^{1.77}$.}
\end{figure}

\begin{deluxetable*}{lrrrrrl} 
\tablenum{7}
\tablecolumns{7}
\tablewidth{0pt}
\tablecaption{Silicate emission analysis\label{Tbl11}}
\tablehead{
  \colhead{ } & \colhead{ } & \colhead{Dust Em.} & \colhead{ } & 
  \colhead{ } & \colhead{Corrected} & \colhead{Infrared} \\
  \colhead{Target} & \colhead{[7]$-$[15]} & \colhead{Contrast} & 
  \colhead{$F_{10}/F_{11}$} & \colhead{$F_{10}/F_{12}$} & 
  \colhead{$F_{10}/F_{11}$} & \colhead{Sp. Class}
}
\startdata
BFM 1             & 0.41$\pm$0.07 & 0.24$\pm$0.04 & \nodata       & \nodata       & \nodata       & 2.ST\\
BFM 1 (follow-up) & 0.62$\pm$0.06 & 0.16$\pm$0.02 & \nodata       & \nodata       & \nodata       & 2.NO\\
HV 12149          & 0.88$\pm$0.02 & 0.80$\pm$0.04 & 1.39$\pm$0.05 & 2.29$\pm$0.13 & 1.68$\pm$0.11 & SE8\\
Massey SMC 59803  & 0.58$\pm$0.01 & 0.23$\pm$0.01 & 0.77$\pm$0.03 & 1.08$\pm$0.04 & 1.19$\pm$0.06 & SE4\\
\\
IRAS 04509        & 2.12$\pm$0.01 & 3.96$\pm$0.05 & 1.17$\pm$0.02 & 2.06$\pm$0.07 & 1.59$\pm$0.06 & SE8\\
IRAS 04516        & 2.00$\pm$0.01 & 3.22$\pm$0.02 & 1.02$\pm$0.01 & 1.47$\pm$0.04 & 1.38$\pm$0.04 & SE6\\
IRAS 04545        & 2.22$\pm$0.01 & 1.51$\pm$0.01 & \nodata       & \nodata       & \nodata       & 3.SA\\
HV 888            & 1.60$\pm$0.01 & 1.90$\pm$0.01 & 1.27$\pm$0.02 & 2.00$\pm$0.05 & 1.58$\pm$0.05 & SE8\\
HV 2310           & 0.88$\pm$0.03 & 0.63$\pm$0.02 & 1.00$\pm$0.04 & 1.26$\pm$0.02 & 1.30$\pm$0.05 & SE5\\
IRAS 05128        & 1.93$\pm$0.00 & 2.42$\pm$0.02 & 0.95$\pm$0.01 & 1.25$\pm$0.02 & 1.29$\pm$0.03 & SE5\\
HV 5810           & 1.02$\pm$0.02 & 0.19$\pm$0.08 & \nodata       & \nodata       & \nodata       & 2.U\\
WOH G 339         & 0.92$\pm$0.02 & 0.46$\pm$0.02 & 1.11$\pm$0.04 & 1.14$\pm$0.06 & 1.27$\pm$0.08 & SE5\\
WBP 77            & 1.01$\pm$0.10 & 0.19$\pm$0.07 & 0.72$\pm$0.14 & 0.43$\pm$0.10 & 0.88$\pm$0.26 & SE1:\\
HV 2572           & 1.06$\pm$0.01 & 0.80$\pm$0.02 & 1.23$\pm$0.02 & 1.73$\pm$0.03 & 1.49$\pm$0.03 & SE7\\
HV 2578           & 1.05$\pm$0.01 & 0.85$\pm$0.03 & 1.24$\pm$0.03 & 1.85$\pm$0.06 & 1.53$\pm$0.06 & SE7\\
HV 996            & 1.63$\pm$0.01 & 1.88$\pm$0.01 & 1.19$\pm$0.01 & 1.65$\pm$0.04 & 1.46$\pm$0.04 & SE7\\
IRAS 05329        & 1.94$\pm$0.00 & 1.14$\pm$0.03 & \nodata       & \nodata       & \nodata       & 3.SB\\
IRAS 05402        & 2.24$\pm$0.00 & 2.29$\pm$0.04 & \nodata       & \nodata       & \nodata       & 3.SA\\
HV 12667          & 0.92$\pm$0.01 & 0.57$\pm$0.03 & 1.17$\pm$0.02 & 1.54$\pm$0.06 & 1.42$\pm$0.06 & SE6\\
HV 12070          & 0.60$\pm$0.02 & 0.29$\pm$0.02 & 1.16$\pm$0.01 & 1.50$\pm$0.02 & 1.41$\pm$0.03 & SE6\\
\enddata
\end{deluxetable*}

The classification method introduced by \cite{sp95,sp98} 
quickly distinguishes the composition of the emitting 
oxygen-rich dust.  We apply that method here, fitting an 
assumed photospheric spectrum to the blue end of the observed
spectrum and subtracting it to isolate the dust contribution.  
Because the photospheric spectra in the Magellanic Clouds 
have weaker SiO absorption than in the Galactic sample (see 
\S 4), we depart from the previous method and simply assume a 
3600 K Planck function for the star.  We fit it to the 
spectrum in the interval 6.8--7.4~\mum\ and subtract it.  The 
classification is based on the ratios of the flux from the 
dust spectrum in small wavelength intervals centered at 10, 
11, and 12~\mum.  Figure 11 shows that the flux ratios from 
oxygen-rich dust trace the {\it silicate dust sequence}.  We 
classify the spectra based on their position up this 
sequence, starting with SE1 in the lower left and running to 
SE8 in the upper right.  The silicate emission index = 
$10 (F_{11}/F_{12}) - 7.5$, where $F_{11}/F_{12}$ is the 
corrected ratio found by locating the point on the power law 
$F_{10}/F_{12} = 1.32 (F_{10}/F_{11})^{1.77}$ closest to the 
actual value for a given spectrum.  Table 7 presents the 
results.

We also measured the magnitude  of the stars in two narrow
filters (6.8--7.4 and 14.4--15.0~\mum) to produce a [7]$-$[15]
color.  Table 7 includes these colors; we will examine these 
colors more closely when comparing other samples in a future 
paper.



\subsection{Crystalline silicate structure beyond 20~\mum} 

\begin{figure} 
\includegraphics[width=3.5in]{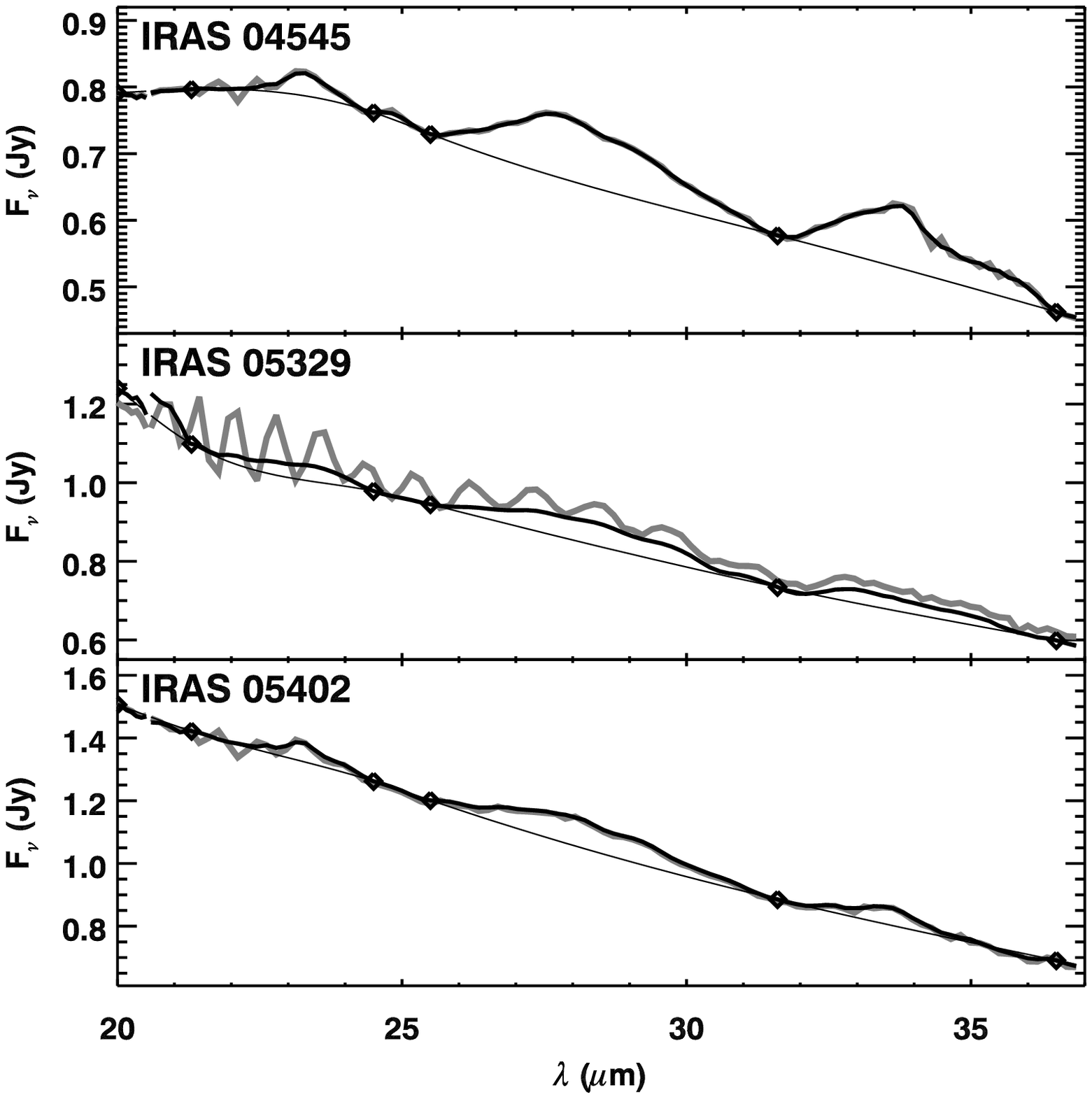}
\caption{The crystalline silicate emission features apparent
in three LL spectra.  The data plotted in gray are the 
original spectra, while the data plotted in black are the
spectra which have been defringed by boxcar smoothing, as
explained in the text.  The thin black line is the spline
fitted at wavelengths marked by diamonds.  Fig.\ 13 presents
the resulting crystalline silicate spectra.}
\end{figure}

\begin{figure} 
\includegraphics[width=3.5in]{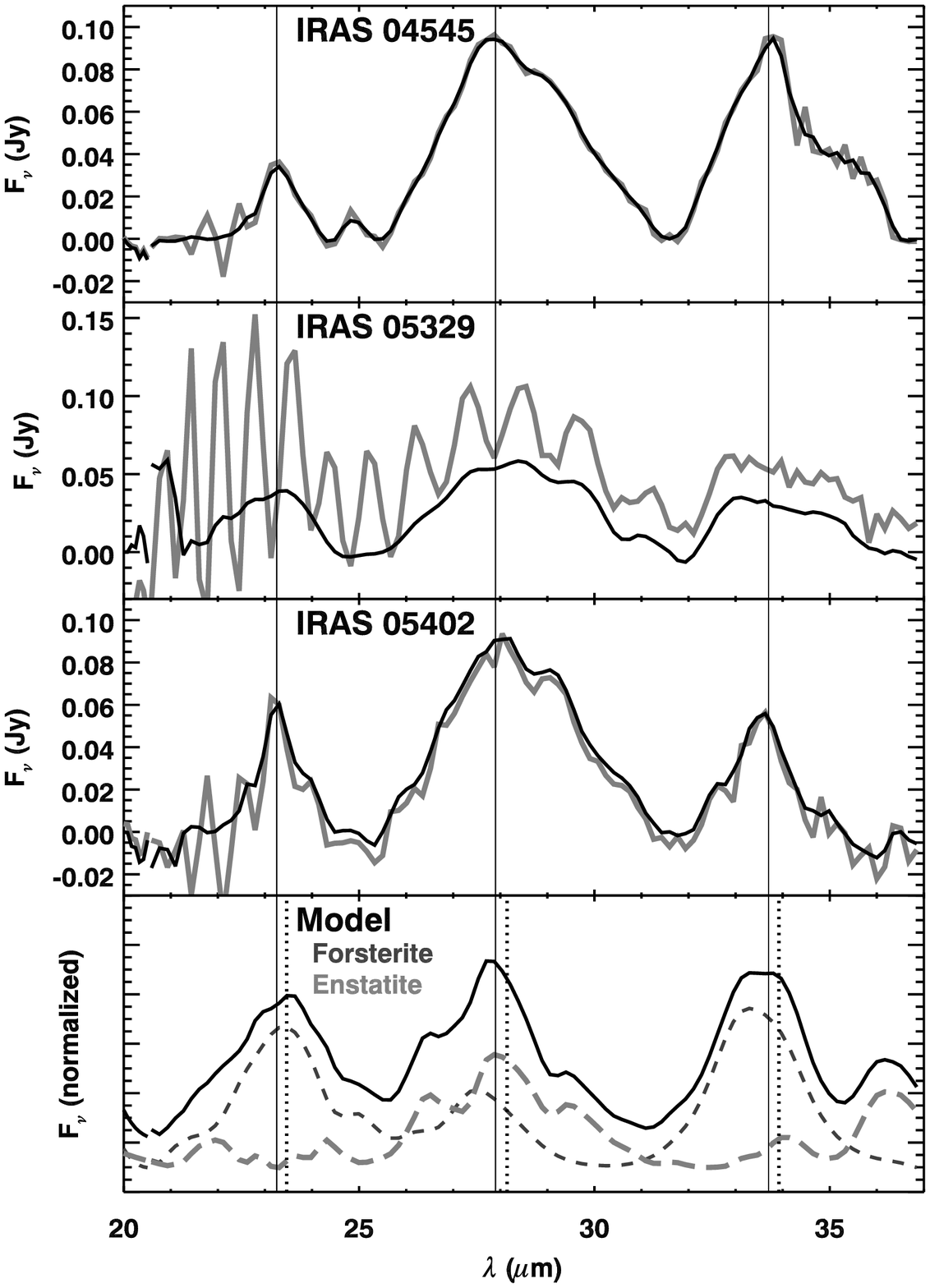}
\caption{The spline-subtracted crystalline silicate emission
features from the spectra in Fig.\ 12.  In the top three 
panels, the smoothed data are in black and the unsmoothed 
data are in light gray.  The vertical lines mark reference 
wavelengths at 23.25, 27.9, and 33.7~\mum.  The bottom panel 
presents a model spectrum (black) composed of 50\% laboratory
forsterite \citep[Mg$_2$SiO$_4$, dark gray;][]{fab01b} and 50\% 
laboratory enstatite \citep[MgSiO$_3$, light gray;][]{jag98},
both at 80 K and assuming a CDE2 shape distribution (see 
text).  The vertical dotted lines in the bottom panel 
illustrate how the observed peaks should shift from 80 K
to 300 K, based on the laboratory work by \cite{sut06}.}
\end{figure}

Figure 10 shows the spectra of the three most embedded 
sources in the MC\_DUST sample, all in the LMC.  Two of these 
sources show clear evidence of crystalline silicate emission 
features at 23, 28, and 34~\mum.  The third, IRAS 05329, 
shows strong fringing in LL1 that obscures the crystalline 
feature at 23~\mum\ and partly hides the one at 28~\mum.  
These crystalline features are much broader than the 
instrumental resolution, so a boxcar smoothing algorithm
applied to LL1 will mask the fringes without significantly
degrading the emission features.  For IRAS 04545 and 
IRAS 05402, we applied a 3-pixel boxcar to all wavelengths 
and then applied an additional 5-pixel boxcar to the LL1 data
to the blue of 22.88~\mum.  For IRAS 05329, with its stronger 
fringing pattern, we applied a 5-pixel boxcar at all 
wavelengths and a second 5-pixel boxcar to the blue of 
30.5~\mum.  We stitched and trimmed these defringed spectra 
in the same manner as described in \S 3.1.  Figure 12 
presents the results.  (Correcting the fringes changed the 
segment-to-segment normalization.)


We fitted a spline to the data at 20.0, 21.3, 24.5, 25.5, 
31.6, and 36.5~\mum\ to produce the spectra in Figure 13.
The bottom panel in Figure 13 plots a model constructed
from laboratory spectra of forsterite \citep{fab01b} and 
enstatite \citep{jag98}, the magnesium-rich endmembers of the 
olivine (Mg$_2$SiO$_4$) and pyroxene (MgSiO$_3$) series, 
respectively.  We have produced optical efficiencies from the
published indices of refraction for a modified continuous 
distribution of ellipsoids which uses quadratic weighting to 
emphasize spherical grains (CDE2), following the methods
described by \cite{fab01b}.  We found that a grain temperature
of 80 K reproduced the average relative strengths of the 23 
and 34~\mum\ features, which both arise from forsterite.
In order to reproduce the strong emission feature at 28~\mum,
our sample had to include equal parts enstatite and 
forsterite.  Adding enstatite also improves the match to the 
wavelength of the observed feature at 28~\mum.


Spectra from the SWS on \iso\ of Galactic sources of
crystalline silicates serve as a useful comparison.
\cite{mol02} found important differences between spectra
produced by objects with known disks and objects where the
emitting dust was in stellar outflows, particularly at 28 and 
34~\mum.  Their Figures 3, 4, and 5 present mean spectra from
disk and outflow objects for the 23, 28, and 34~\mum\ 
features, respectively.  For comparison, we concentrate on 
IRAS 04545 and IRAS 05402 due to the higher quality of their 
spectra compared to IRAS 05239.  In both spectra, the 
23~\mum\ feature shows a sharp peak at 23.25~\mum.  This 
differs from both the Galactic disk sources, which show a 
peak at 23.6~\mum, and the Galactic outflow sources, which 
have a broader peak from 22.8 to 23.8~\mum.  The 28 and 
34~\mum\ features in the LMC sources both resemble the 
Galactic disk sources.  In the LMC sources, the 28~\mum\ 
feature peaks at 27.9~\mum\ and is as strong as the 23 and
34~\mum\ features.  In the Galactic samples, the 28~\mum\
feature is largely absent in the outflow sources, but 
strong in the disk sources.\footnote{The spectra produced 
by \cite{mol02} show additional structure at $\sim$29.5 and 
30.5~\mum, but this spectral region is notoriously difficult 
in the SWS data, due to problems in spectral bands 3D and 3E, 
making this structure questionable \citep[][discuss this in 
more detail]{slo03}.}  In the LMC sources, the 34~\mum\ 
feature peaks at 33.7~\mum, as does the Galactic disk sample, 
while the Galactic outflow sample shows an additional peak at 
32.9~\mum\ which appears as only a shoulder in the disk
spectra.  Thus, while the 23~\mum\ feature differs somewhat
from the Galactic sample, both the 28 and 34~\mum\ features
closely resemble spectra from Galactic sources of crystalline
silicates, especially those found in circumstellar disks.

Several effects can shift the wavelength of the observed
features.  \cite{sut06} recently investigated how the emission 
features from forsterite shift with temperature.  Using their 
data, we could expect a shift to the red of 0.21~\mum\ in the 
23.25~\mum\ feature, 0.25~\mum\ at 27.9~\mum, and 0.22~\mum\ 
at 33.7~\mum\ if we were to heat the grains in our observed 
spectra from 80 K to 300 K, which is the temperature of the 
measured laboratory samples.  The bottom panel of Figure 13 
includes vertical dotted lines illustrating these shifts.

Another way to shift the features is to replace some of the
Mg in the minerals with Fe.  \cite{chi02} and \cite{koi03}
studied this effect in pyroxenes and olivines, respectively.
In pyroxenes, replacing 20\% of the Mg with Fe shifts the 
28~\mum\ feature by 0.4~\mum, while a similar replacement in 
olivines shifts the 23~\mum\ feature 0.3~\mum\ and the 
34~\mum\ feature 0.2~\mum.  

The shape of the grains also influences the wavelengths of
the emission features, as \cite{min03} have thoroughly
investigated.  Generally, a shift from spherical to
non-spherical shape distributions moves the features to
longer wavelengths.  Most examinations of this effect have
assumed that no particular axis is favorably extended in a
sample of grains.  \cite{slo06a} examined the effect of
preferential non-sphericity along different axes and found
that some shapes could shift the features to the blue, 
instead of the red.

The solid vertical lines in Figure 13 mark the average
peaks of the features in IRAS 04545 and IRAS 05402.  In
the bottom panel, we should compare the peaks of the
model spectrum to the {\it dotted} vertical lines, since
these account for the difference in temperature between
the laboratory and LMC data.  The observed position of the 
23~\mum\ forsterite feature shows no redshift compared to 
the laboratory data, indicating Mg-rich grains with a CDE2 
shape distribution are a good fit. The 28~\mum\ feature, 
which arises from a mixture of enstatite and forsterite, is 
$\sim$0.3~\mum\ to the red of the laboratory data, indicating 
Fe inclusions of no more than $\sim$15\% or a slightly 
different shape distribution.  Less can be said about the
relative positions of the 34~\mum\ feature in the observed 
and laboratory data because of the clear difference in 
profile.  Overall, our observations of crystalline silicates 
in the LMC are fully consistent with Mg-dominated grains, 
much like those observed in the Galaxy.

\subsection{Crystalline silicate structure at 10~\mum} 

\begin{figure} 
\includegraphics[width=3.5in]{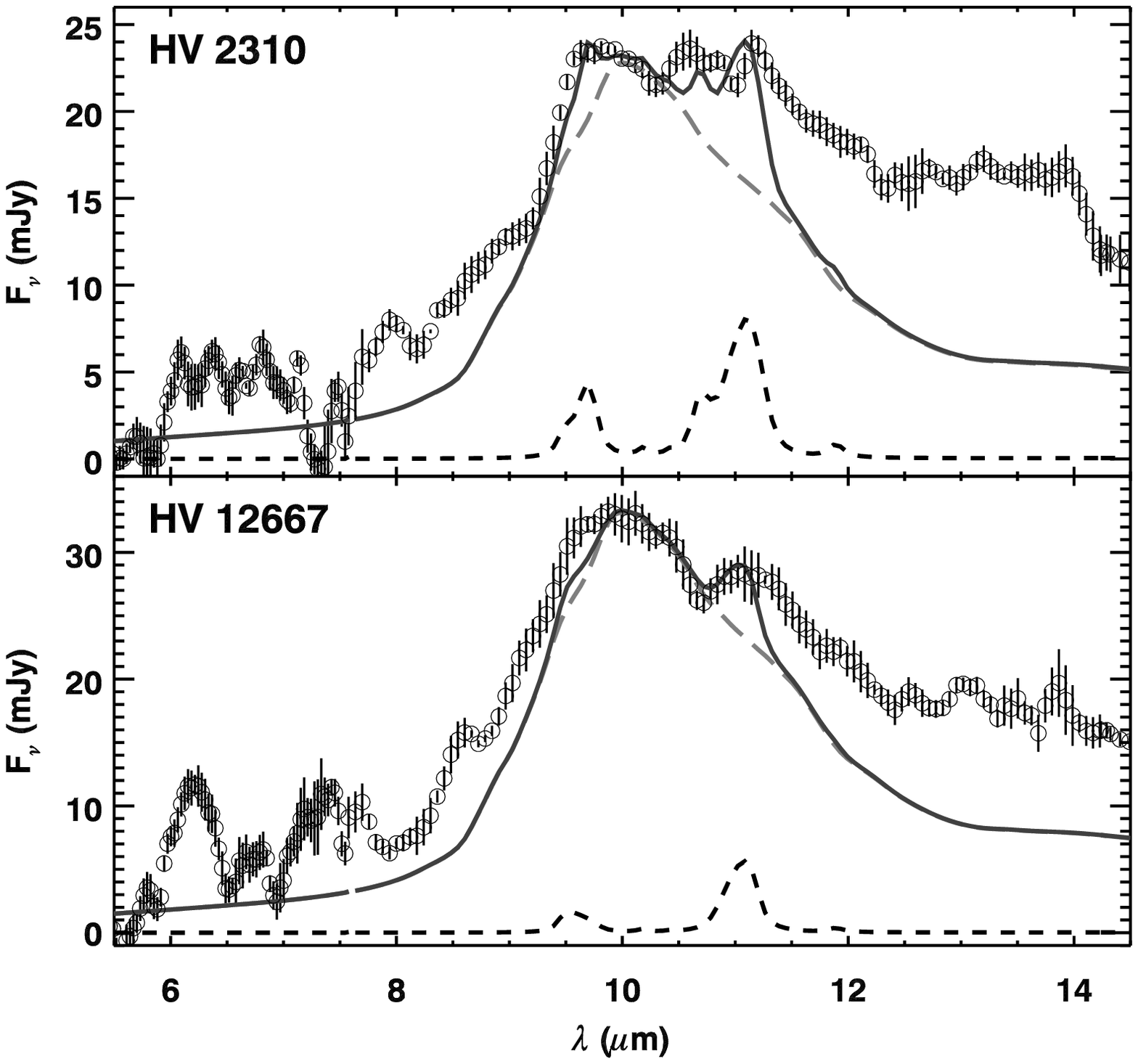}
\caption{Structured silicate emission at 10~\mum\ in HV 2310
(top) and HV 12667 (bottom).  The IRS data (circles and error
bars) have been smoothed with a 3-pixel boxcar.  The fitted
model dust spectrum (solid black line) is a combination of 
amorphous silicates \citep[dashed gray curve;][]{ohm92} and 
forsterite \citep[Mg$_2$SiO$_4$; dashed black curve;][]{fab01b}. 
For HV 2310, 7\% of the grains are crystalline, with a shape
distribution including spherical, prolate (along the $x$ 
axis), and oblate (extended along the $x$ and $z$ axes).  For 
HV 12667, 2.5\% of the dust is crystalline, and the shape 
distribution includes spherical grains and prolate grains 
($x$ axis).}
\end{figure}

\cite{slo06a} published the spectrum of HV 2310, one of the
first objects observed in the MC\_DUST program.  This 600-day
Mira shows an unusual silicate emission spectrum with 
structure in the 10--11~\mum\ region best fit with a mixture
of 7\% crystalline forsterite and a blend of amorphous grains
dominated by astronomical silicates \citep[as described 
by][]{ohm92}.  They based their analysis on S11 pipeline
output from the SSC, while the data here are S15.  While the 
general shape is still indicative of crystalline silicates,
the spectral details have changed enough to require a new
analysis.

The MC\_DUST sample has revealed one more object showing
crystalline silicate structure at 10~\mum, HV 12667.  This 
source is a 645-day Mira, so it has similar pulsation 
properties to HV 2310, and it also has a similar infrared
brightness, 45 mJy at 8~\mum, compared to 36 mJy for HV 2310.

Figure 14 plots the continuum-subtracted spectra of HV 2310
and HV 12667.  In both cases, we have fitted and subtracted
a 3600 K Planck function at 5.3--5.8~\mum.  We tested the
dependency of our results on the assumed stellar continuum
by also using a model M6 giant with SiO absorption at 8~\mum,
and we found no significant impact.  Figure 14 also includes 
a rough fit to the dust spectrum, using a combination of 
amorphous silicates \citep{ohm92} and forsterite 
\citep{jag98}.  Following \cite{slo06a}, we have produced a 
modified shape distribution in an effort to reproduce the 
detailed structure of the 10~\mum\ silicate feature.  In the 
case of HV 12667, we have used a distribution of spherical 
grains, and prolate grains extended along the $x$ axis.  To 
fit the observed spectrum, this population of crystalline 
grains needs to account for 2.5\% of the dust; the rest is 
amorphous silicates.  For HV 2310, the shape distribution 
includes prolate grains ($x$ axis) and oblate grains ($x$ and 
$z$ axes).  The crystalline component of the dust population 
is 7\%, unchanged from before.  \cite{slo06a} also included 
amorphous pyroxene and amorphous alumina to fit the blue and 
red wings of the silicate profile, respectively.  We do not 
include these components because we wish to concentrate on 
the structure at the peak of the profile.


It is highly unlikely that these shape distributions fit the
silicate profiles uniquely.  In fact, the changes in the 
pipeline calibration have changed the spectrum of HV 2310
enough that we have modified the shape distribution used by
\cite{slo06a}.  The prolate grains in their shape 
distribution were extended along the $z$ axis, vs.\ the $x$ 
axis here.  This difference illustrates that the data are not
of sufficient fidelity to pin these details down.  However,
the fits to HV 2310 and HV 12667 demonstrate that the
silicate profiles in both are consistent with forsterite and 
that is is not necessary to invoke enstatite or minerals with 
Fe components to fit the features.  

\subsection{The 13~\mum\ emission feature} 

The 13~\mum\ emission feature, first detected by 
\cite{lmp86}, often appears in the spectra from Galactic
oxygen-rich dust shells \citep{slo96}.  In our Magellanic
samples, it is largely absent.  There may be a feature in
the spectra of HV 2572 and HV 2578, but if so, it is weak,
noisy, and difficult to measure.  The spectra of IRAS 04509, 
IRAS 04516, and IRAS 05128 show features in this spectral 
region, but they are centered closer to 14~\mum, as discussed 
in the next section.

Our sample of oxygen-rich dust spectra is small, it may 
contain some biases (e.g.\ for Mira variables, which are 
less likely to show a 13~\mum\ feature than semiregular 
variables), and some of the spectra are noisy, which can hide 
low-contrast emission features.  Consequently, our conclusion
that the prevalence and strength of the 13~\mum\ emission 
feature is reduced in the Magellanic Clouds is somewhat
preliminary.  A more thorough examination of oxygen-rich 
circumstellar dust observed in other Magellanic programs is 
in preparation.

\subsection{The 14~\mum\ emission feature} 

\begin{figure} 
\includegraphics[width=3.5in]{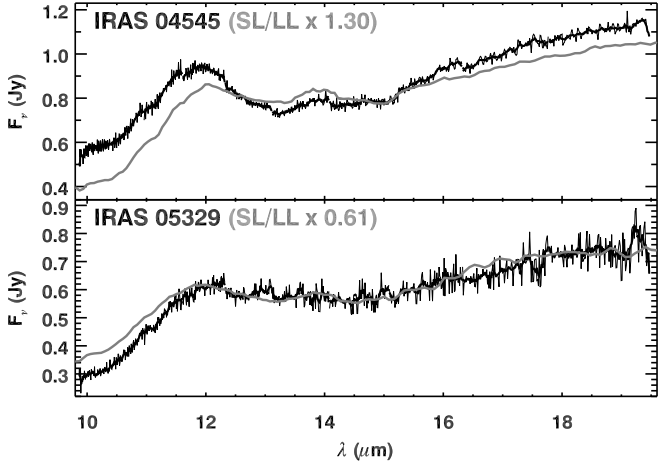}
\caption{The 14~\mum\ feature in IRAS 04545 (top) and 
IRAS 05239 (bottom), in low-resolution IRS data (light
gray) and SH data.  The SH spectra are plotted before and 
after smoothing with a 7-pixel boxcar.
IRAS 04545 increased 30\% in brightness,
producing a high-quality SH spectrum which confirms the
presence of the 14~\mum\ feature.  IRAS 05329, on the other
hand, dimmed by 40\%, making confirmation more difficult.}
\end{figure}

\cite{slo06a} previously noted that the spectrum of HV 2310 
includes an emission feature at 14~\mum.  They were somewhat 
cautious about the feature, since it appears at the boundary 
between SL and LL, raising the possibility that it is an 
artifact, even if no known effect in the IRS could produce 
such a feature.  \cite{ec06} also detected a 14~\mum\ 
emission feature in the IRS spectra of the Trojan asteroids 
624 Hektor and 921 Agamemnon, but they were also cautious and
did not suggest a possible carrier.  Figures 9--10 reveal 
that several sources in the MC\_DUST sample besides HV 2310 
also have 14~\mum\ features.  

To test if the feature might be an artifact in the 
low-resolution IRS spectra, we observed two 14~\mum\ sources,
IRAS 04545 and IRAS 05329, with the SH module, as shown in 
Figure 15.   Both sources are variable stars, and IRAS 04545
increased in brightness by 30\% between the two observations,
resulting in an excellent SH spectrum which clearly confirms
the detection in SL and LL.  IRAS 05329 declined in 
brightness by nearly 40\%, and as a consequence, the SH 
spectrum is noisier.  While the SH data are consistent with
the low-resolution data, they do not show an obvious 14~\mum\
feature.  



\subsection{BFM 1} 

\begin{figure} 
\includegraphics[width=3.5in]{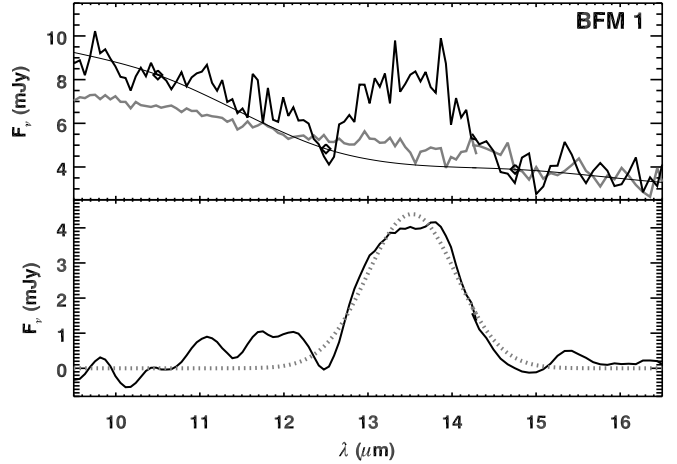}
\caption{The spectrum of BFM 1 from 9.5 to 16.5~\mum, as 
originally observed (in black), and when followed up with 
a longer integration time (in gray) 418 days later.  The 
thin black curve is a spline fitted through the original
spectrum to isolate the possible dust spectrum shown in the
bottom panel.  In the bottom panel, the spectrum has
been smoothed, and the fitted Gaussian 
(dashed gray curve) is centered at 13.52~\mum\ and has a 
FWHM of 0.53~\mum.} 
\end{figure}

As shown in Figures 6, 7, and 8, BFM 1 has some unusual 
spectral characteristics.  First, while the spectrum shows no
substantial structure out to 12.5~\mum, it has a significant 
dust emission contrast (0.24 when first observed, and 0.16 
when followed up), indicating the presence of a featureless
excess component.  Second, the original spectrum shows a
previously unknown emission feature from 13 to 14~\mum.
Third, the follow-up spectrum, taken 418 days later with a
longer integration time, confirms the featureless dust 
excess, but shows nothing where the 13--14~\mum\ feature
had previously been.

BFM 1 is the only true S star in our sample, and it is a
steady Mira variable with a 398-day period and an amplitude
of about 4 magnitudes in the optical.  In S stars, the 
C/O ratio is near unity, leaving little C or O for the 
production of dust once CO has formed, which could lead to
unusual dust chemistry.  Our spectra of BFM 1 certainly
show unusual characteristics.  The featureless excess is
not consistent with amorphous alumina, as that would leave
a recognizable low-contrast feature in the 11--12~\mum\
region.  Perhaps it is amorphous carbon, or some other
dust component.

The possible feature at 13--14~\mum\ is also enigmatic.
The follow-up spectrum did not confirm its presence, so
either it disappeared in the interval between the 
observations, or else it was an artifact to begin with.
The two spectra were obtained 418 days apart, compared to
the 398-day period of the star.  Thus the spectra were
at nearly the same phase, ruling out some phase-dependent 
phenomenon.  Nonetheless, we suspect the feature is real.  
First, it appears in the spectra from both nod positions.  
Second, it covers over 25 pixels in SL and is not the 
product of any known artifact.  Third, similar features 
appear in two of the 90 spectra in the sample of Galactic 
S stars observed by the IRS in Cycle 3 \citep[Program 
ID 30737, P.I.\ S.\ Hony]{slo08b}.

Figure 16 focuses on the 13--14~\mum\ feature in BFM 1.
Fitting a spline to remove the continuum isolates a feature 
which can be fit with a Gaussian centered at 13.52~\mum\ 
with a full width at half maximum (FWHM) of 0.53~\mum.
The unique spectrum of BFM 1 deserves a unique 
classification.  We extend the Hanscom system with a 
new classification, 2.ST, where the ``T'' indicates a
feature centered at 13.5~\mum.


\section{Carbon-rich Sources} 

\begin{figure} 
\includegraphics[width=3.5in]{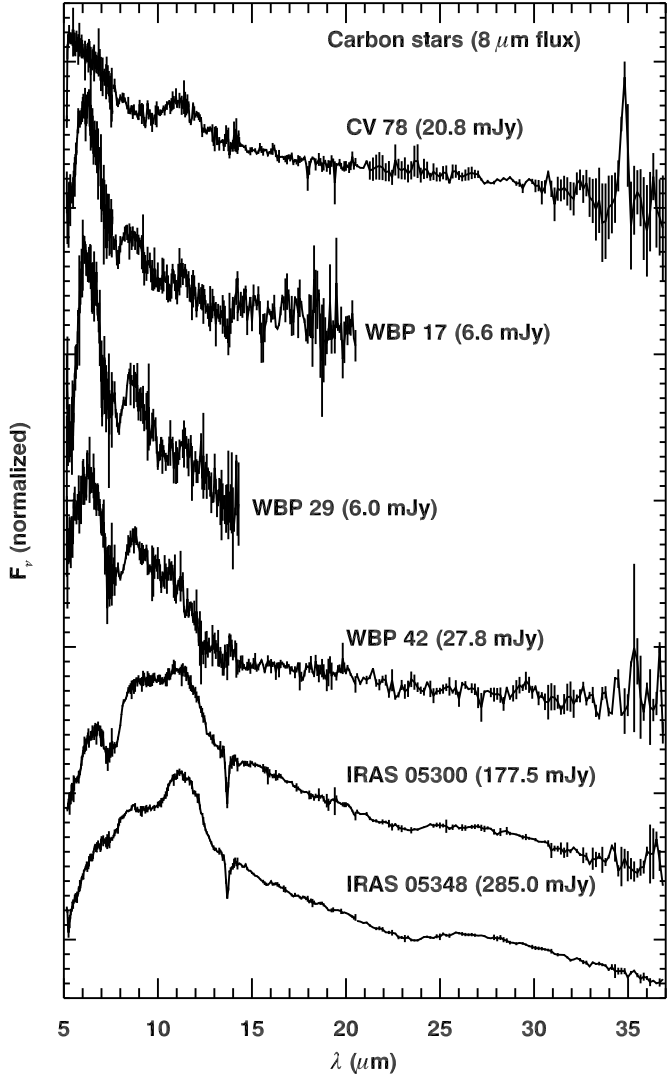}
\caption{The six definite carbon stars in the MC\_DUST 
sample.  CV 78, the top spectrum, is in the SMC; the rest 
are in the LMC.  The spectra of WBP 17 and WBP 29 are
truncated at longer wavelengths due to background issues.
The spectrum of a seventh source, MACHO 79.5505.26, appears 
in Fig.\ 2 and 4 and has an uncertain chemistry.}
\end{figure}

The MC\_DUST sample includes one definite carbon star in the 
SMC and five in the LMC.  Figure 17 presents the spectra of 
these sources.  In addition, MACHO 79.5505.26 has a 
carbon-rich optical spectrum but an ambiguous infrared 
spectrum with no dust emission and only CO absorption at
5~\mum.  Its spectrum appears in Figures 2 and 4,   In the
next section, we show that it has no carbon-rich absorption
features.  We exclude it from further analysis as a carbon 
star because the chemistry of the infrared component
is uncertain.  

We have truncated the spectra of two sources in Figure 17
to avoid some problems introduced by complex backgrounds.
The LL1 data for WBP 17 are negative due to strong 
background gradients.  A nearby bright carbon star 
contaminates the LL data for WBP 29.


\subsection{The Manchester Method} 

\begin{deluxetable*}{lcccccccl} 
\tablenum{8}
\tablecolumns{9}
\tablewidth{0pt}
\small
\tablecaption{Carbon star analysis\label{Tbl12}}
\tablehead{
  \colhead{ } & \colhead{[6.4]$-$[9.3]} & \colhead{[16.5]$-$[21.5]} & 
  \colhead{7.5~\mum\ EW} & \colhead{13.7~\mum\ EW} &
  \colhead{$\lambda_{SiC}$} & \colhead{ } & \colhead{ } & 
  \colhead{Infrared} \\
  \colhead{Target} & \colhead{(mag)} & \colhead{(mag)} & \colhead{(\mum)} &
  \colhead{(\mum)} & \colhead{(\mum)} & \colhead{SiC/Cont.} & 
  \colhead{MgS/Cont.} & \colhead{Sp. Class}
}
\startdata
CV 78            &    0.204$\pm$0.048 &    0.178$\pm$0.061 & 0.030$\pm$0.030 & 0.024$\pm$0.034 & 11.12$\pm$0.15 & 0.197$\pm$0.019 & \nodata         & 2.CE \\
MACHO 79.5505.26 &    0.208$\pm$0.044 & $-$0.307$\pm$0.350 & \nodata         & \nodata         & \nodata        & \nodata         & \nodata         & 1.N  \\
WBP 17           & $-$0.113$\pm$0.047 & $-$1.027$\pm$0.770 & 0.304$\pm$0.014 & 0.157$\pm$0.032 & 11.43$\pm$0.39 & 0.092$\pm$0.031 & \nodata         & 1.NC \\
WBP 29           &    0.084$\pm$0.030 &    \nodata         & 0.395$\pm$0.011 & \nodata         & 12.01$\pm$0.69 & 0.035$\pm$0.037 & \nodata         & 2.CE \\
WBP 42           &    0.358$\pm$0.017 &    0.140$\pm$0.094 & 0.310$\pm$0.019 & \nodata         & 10.95$\pm$0.31 & 0.092$\pm$0.025 & \nodata         & 2.CE: \\
IRAS 05300       &    1.061$\pm$0.006 &    0.280$\pm$0.011 & 0.152$\pm$0.003 & 0.047$\pm$0.010 & 11.32$\pm$0.08 & 0.098$\pm$0.005 & 0.255$\pm$0.014 & 3.CR \\
IRAS 05348       &    1.050$\pm$0.006 &    0.295$\pm$0.010 & 0.013$\pm$0.004 & 0.047$\pm$0.001 & 11.28$\pm$0.05 & 0.137$\pm$0.004 & 0.238$\pm$0.012 & 3.CR \\
\enddata
\end{deluxetable*}

To analyze the carbon-rich spectra, we applied the Manchester 
method introduced by \cite{slo06b} and also applied by 
\cite{zij06} and \cite{lag07}.  These papers examined  a 
total of 28 carbon stars in the LMC and 33 in the SMC, while 
our sample adds five carbon stars in the LMC and one in the 
SMC.  The Manchester method measures the spectral continuum 
at four narrow wavelength intervals relatively free of 
molecular band absorption or dust emission centered at 6.3, 
9.4, 16.5, and 21.5~\mum.  The first two wavelengths define 
the [6.4]$-$[9.3] color, which measures the contribution from 
warm dust relative to the stellar photosphere.  The 
[16.5]$-$[21.5] color measures the temperature of any cool 
dust contribution to the spectrum.  It also provides a 
means of estimating the contribution under the MgS dust 
emission feature, which peaks in the 26--30~\mum\ range.  

Gaseous acetylene (C$_2$H$_2$) produces two absorption
bands visible in the IRS spectra:  a broad band centered at
7.5~\mum\ and typically $\sim$1.0~\mum\ wide, and a narrow
band at 13.7~\mum\ \citep{mat06}.  This latter band is 
actually the Q branch of the fundamental $\nu_5$ mode and 
it is often centered in a broader band which includes the P 
and R branches, but the broad component can be difficult to 
recognize due to the presence of the SiC dust emission 
feature at 11.3~\mum.

The Manchester method fits line segments to the continuum to 
measure the equivalent width of the acetylene bands and the
integrated flux of the SiC emission feature.  We express the 
flux of the SiC band as a ratio to the continuum underneath.  
To measure the strength of the MgS the Manchester method uses 
the [16.5]$-$[21.5] color to extrapolate a cool blackbody 
continuum beneath the feature, which peaks in the 
26--30~\mum\ range.  Table 8 presents the results.  


\subsection{Results} 

\begin{figure} 
\includegraphics[width=3.5in]{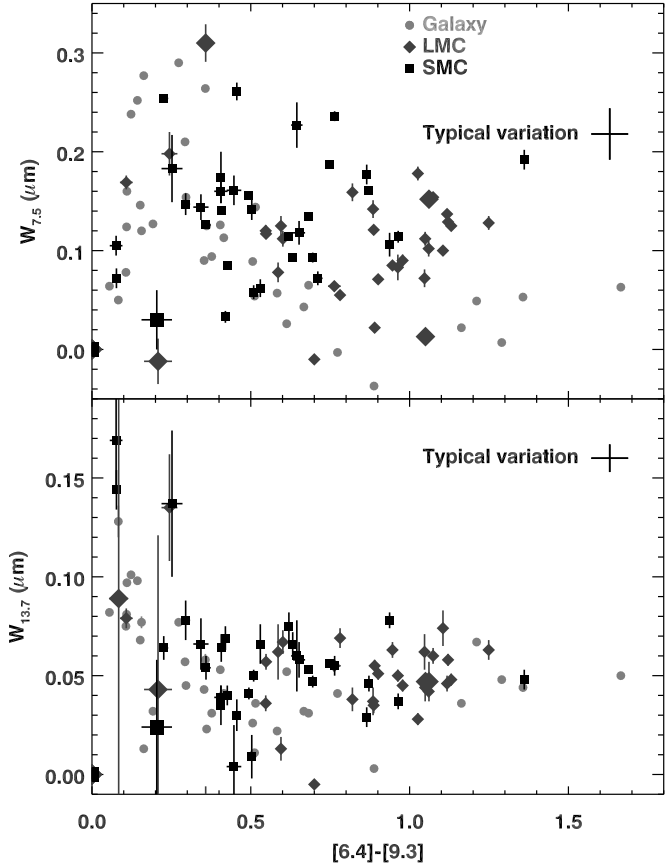}
\caption{The equivalent widths of the acetylene absorption 
bands at 7.5~\mum\ (top) and 13.7~\mum\ (bottom), plotted vs.
[6.4]$-$[9.3] color, which measures the strength of the 
amorphous carbon component.  Galactic data are plotted with
light gray circles, LMC data in dark gray diamonds, and SMC
data in black squares.  The MC\_DUST sample are plotted
with large symbols, while small symbols are from previously
published samples.}
\end{figure}

\begin{figure} 
\includegraphics[width=3.5in]{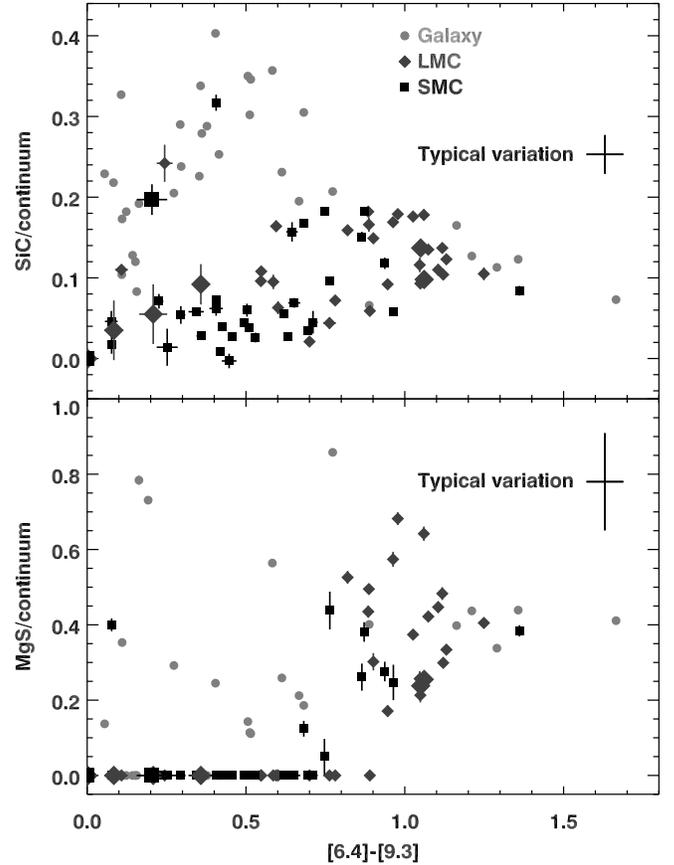}
\caption{The integrated fluxes of the SiC (top) and MgS 
(bottom) dust features, divided by the underlying continuum 
and plotted as a function of [6.4]$-$[9.3] color.  Symbols 
are as defined in Fig.\ 18.}
\end{figure}

Figures 18 and 19 plot the results for our sample (large
symbols), along with the LMC sample of \cite{zij06}, the SMC 
samples of \cite{slo06b} and \cite{lag07}, and the Galactic 
sample observed by the SWS on \iso.  The Galactic sample is 
defined by the carbon-rich classifications of \cite{kra02} 
and uses the data in the SWS Atlas \citep{slo03}.  The small 
MC\_DUST sample follows the trends set by the larger samples 
of Magellanic carbon stars previously published.  As 
metallicity decreases, stars with similar [6.3]$-$[9.4] 
colors show increasing absorption from acetylene and reduced 
emission from SiC and MgS dust.


The strength of the 7.5~\mum\ acetylene band increases as the 
stars go from a [6.4]$-$[9.3] color of 0.0 to $\sim$0.3 and 
then decreases as the amorphous carbon component becomes more
dominant, although the scatter at a given color is large.  
The data appear to follow two separate tracks, but the
significance is not clear with these sample sizes.  The
strength of the 13.7~\mum\ band generally decreases steadily 
from 0.10~\mum\ to $\sim$0.05~\mum\ at a [6.4]$-$[9.3] color 
of 0.5 and remains roughly constant beyond.  In both cases, 
the LMC and SMC sources usually have deeper bands compared to 
the Galactic sources of the same color, but no clear 
difference between the LMC and SMC is discernable.

The strength of the SiC feature follows two distinct tracks 
with [6.4]$-$[9.3] color.  The lower track starts with no SiC
feature at a color of 0.0 and slowly increases to a 
$\sim$10\% strength at a color of 1.0.  The upper track climbs 
to a $\sim$30\% strength at a color of $\sim$0.45, then drops 
to meet the lower track at redder colors.  \cite{slo06b} 
noted that five of their carbon stars in the SMC had enhanced 
SiC strength compared to the rest:  MSX SMC 054, 091, 105, 
159, and 163.  In Figure 19, these five clearly fall on the 
upper track, which includes nearly all of the Galactic 
sources.  \cite{slo06b} were unable to find any other
property that distinguished them from the rest of the SMC
sample, except that they were more likely to show MgS 
emission.  Additional SMC sources on the upper track
include GM 780 from the sample of \cite{lag07} and CV 78
from the MC\_DUST sample.  \cite{lag07} noted that GM 780
was peculiar, with no acetylene absorption and a relatively
high luminosity.  They argued that its C/O ratio was lower
than most of the Magellanic sources and more like Galactic
carbon stars, which might affect the fraction of SiC
in its outflows.

The MgS strength in the SMC and LMC samples are similar
to each other.  Few sources bluer than [6.4]$-$[9.3] $\sim$ 
0.7 having any measurable MgS dust, while most Galactic 
sources do.  At redder colors, the three samples behave 
similarly, with MgS strengths typically $\sim$40\% of the
continuum, but with considerable scatter, especially in the
LMC sample.

\subsection{Distance indicators} 

\begin{figure} 
\includegraphics[width=3.5in]{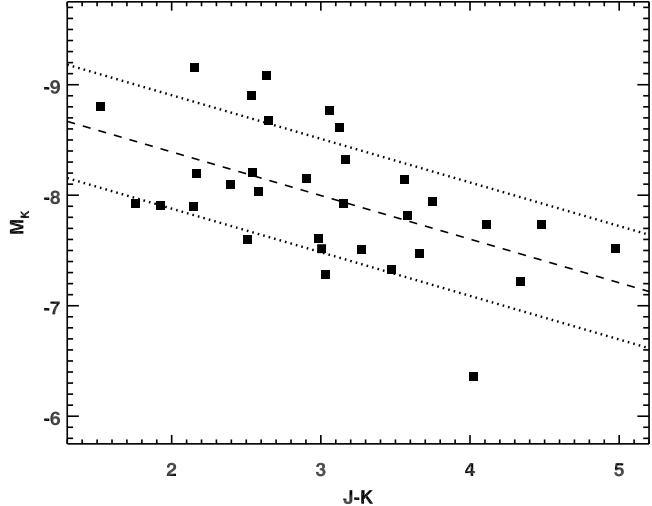}
\caption{A near-infrared color-magnitude diagram for the
carbon stars in the SMC samples of \cite{slo06b} 
and \cite{lag07}.  The dashed line shows
a line fit to the SMC data while the dotted lines show 
the standard deviation of the scatter about that line.}
\end{figure}

\begin{figure} 
\includegraphics[width=3.5in]{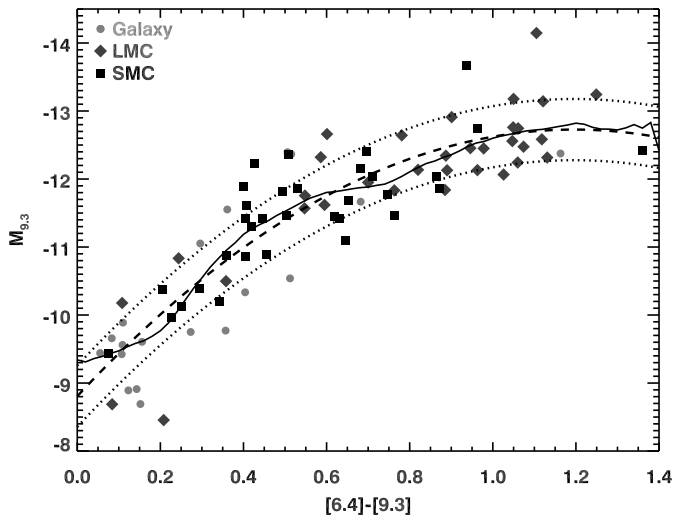}
\caption{A mid-infrared color-magnitude diagram for the
Magellanic and Galactic carbon stars examined by
\cite{slo06b}, \cite{zij06}, and \cite{lag07}.  The
solid line shows the running mean of the data as a function
of [6.4]$-$[9.3] color.  The dashed line is a quadratic fit 
to the data, with the dotted lines above and below showing 
the (smoothed) standard deviation.}
\end{figure}

Figure 20 plots the absolute magnitude at K vs.\ the J$-$K
2MASS colors for the carbon stars in the SMC samples of
\cite{slo06b} and \cite{lag07}.  We have
assumed a distance modulus of 18.93 to the SMC \citep{kw06}.
The data follow a linear relation with a standard deviation
of 0.51 magnitudes:

\begin{equation}
  M_K = -9.18 + 0.395 (J-K).
\end{equation}


Applying the synthetic photometry to the spectra to generate
magnitudes at 6.4 and 9.3~\mum\ reveals a second infrared 
color-magnitude relation.  Figure 21 plots the absolute 
magnitude at 9.3~\mum\ as a function of [6.4]$-$[9.3] color 
for the samples of carbon stars in the Galaxy and both 
Magellanic Clouds of \cite{slo06b}, \cite{zij06}, and 
\cite{lag07}.  The magnitudes at 9.3~\mum\ are based on a 
flux density at zero magnitudes of 45.7 Jy and distance 
moduli to the LMC and SMC of 18.54 and 18.93, respectively 
\citep{kw06}.  The Galactic sample were observed with the 
SWS on \iso, as described by \cite{slo06b}.  The distances 
were estimated by \cite{bc05}.  Here, the relation is 
quadratic:

\begin{equation}
  M_{9.3~\mu m} = \Sigma a_i ([6.4]-[9.3])^i,
\end{equation}

\noindent where $a_0 = -8.81$, $a_1 = -6.56$, and 
$a_2 = 2.74$.  The standard deviation of the data about this 
relation is 0.45 magnitudes.


\section{PAH Emission Sources} 

\begin{figure} 
\includegraphics[width=3.5in]{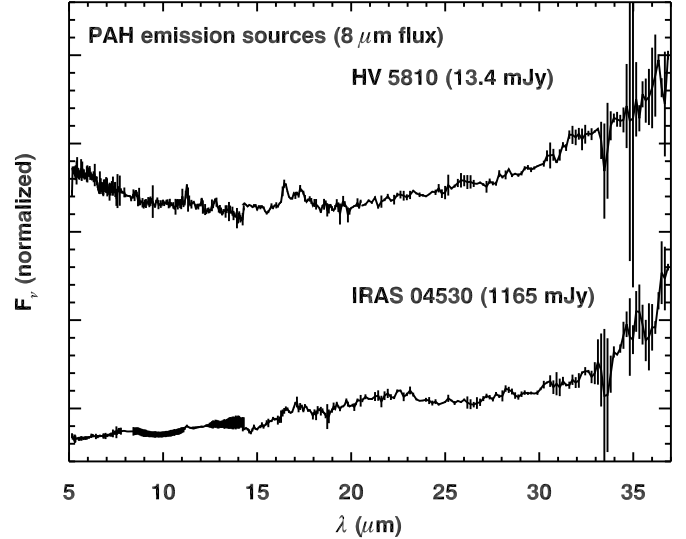}
\caption{Spectra of the two red sources in the MC\_DUST 
sample.  Both show PAH emission features.  A third 
spectrum with PAH features, of Massey SMC 59803, appears in 
Fig.\ 8.}
\end{figure}

\begin{figure} 
\includegraphics[width=3.5in]{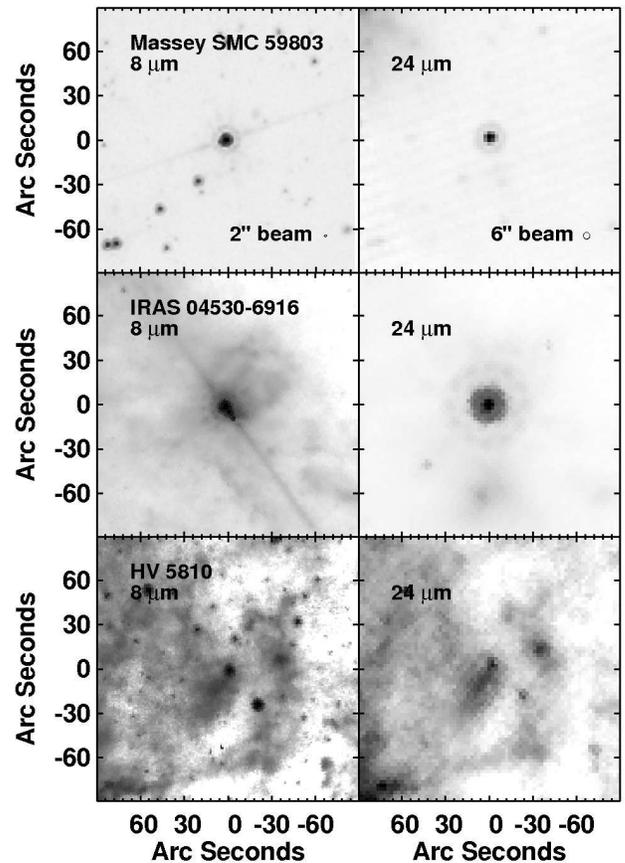}
\caption{The 8 and 24~\mum\ fields of the three PAH sources
shown in Fig.\ 8 and 22.  The images of Massey SMC 59803 and 
IRAS 04530 are plotted on a logarithmic scale, while HV 5810 
is plotted on a linear scale.  Massey SMC 59803 is an 
isolated supergiant, but the other two sources are associated
with extended emission.}
\end{figure}

Figure 22 presents the spectra of the two red sources in the
MC\_DUST sample, IRAS 04530 and HV 5810.  Both of these 
spectra have PAH emission features.  A third source, the
supergiant Massey SMC 59803, also shows PAH emission 
superimposed on an oxygen-rich dust spectrum.  Its spectrum
appears in Figure 8.


Figure 23 includes 8 and 24~\mum\ images of the three PAH
emission sources.  The data for Massey SMC 59803 were 
obtained as part of the S$^3$MC survey by \cite{bol07}.  
Data for the other two sources are from the SAGE survey 
\citep{mei07}.  


\subsection{Massey SMC 59803} 

\begin{figure} 
\includegraphics[width=3.5in]{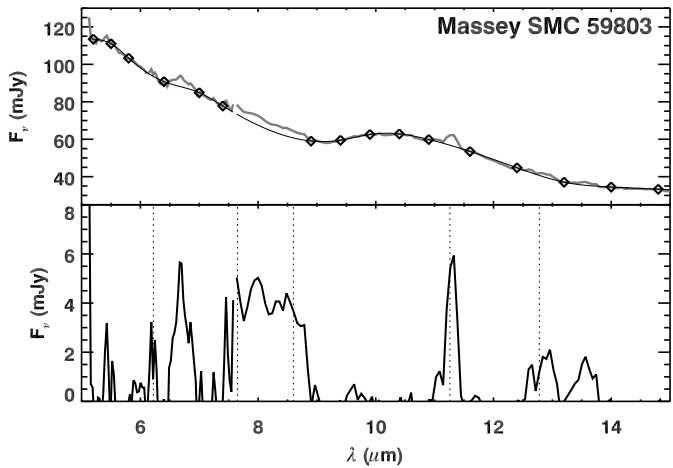}
\caption{The spectrum of Massey SMC 59803 with a fitted
spline ({\it top}), and the residual PAH spectrum ({\it 
bottom}).  The dashed vertical lines in the bottom panel
are at wavelengths of 6.23, 7.70, 8.60, 11.26, an 
12.70~\mum\, the centers of the PAH features in a typical
class A PAH spectrum \citep[in the classification system
introduced by][]{pee02}.}
\end{figure}

An examination of the 8 and 24~\mum\ images of Massey SMC
59803 reveals no associated extended emission.  The IRS data 
are also consistent with a point source at all wavelengths.  
To separate the contribution of PAHs from the spectrum, 
we fit a spline through the ``continuum'' from the star and 
amorphous silicate grains which produce the 10~\mum\ feature, 
as illustrated in Figure 24 (top).  \cite{kel08} applied this 
method to a sample of spectra from Herbig Ae/Be stars with 
both silicate and PAH emission and successfully disentangled 
the separate dust components in most cases.  


Figure 24 (bottom) shows the resulting PAH spectrum.  PAHs in 
the spectra of supergiants are unusual, but not unheard of.  
\cite{syl98} observed PAH emission at 8.6 and 11.3~\mum\ in 
three supergiants in the h and $\chi$ Persei association.
The residual PAH spectrum from Massey SMC 59803 shows 
emission features centered at 6.21, 6.70, 8.13, 11.29, 12.86, 
and 13.56~\mum.  The 6.7~\mum\ feature probably does not 
arise from PAHs and is presently unidentified.  The remaining 
features are from PAHs.  The strongest features are the C--C 
complex centered at 8~\mum\ and the out-of-plane C--H solo 
bending mode at 11.3~\mum.  

\cite{pee02} defined three classes of PAH spectra based 
primarily on the position of the 8~\mum\ PAH complex.  Most 
spectra fell into class A (peak at 7.65~\mum) or class B
(7.85~\mum), but two class C spectra showed a feature 
peaking at $\sim$8.2~\mum.  The detailed position and shape 
of the 8~\mum\ complex depends critically on how the spline 
fits the continuum on the blue side of the silicate feature,
so we should not overinterpret this feature.  Nonetheless,
there does appear to be an excess between $\sim$7.5 and 
9~\mum, and it is more consistent with class C than with
the other classes.  The 11.3~\mum\ feature is also shifted
from its nominal position in class A and B spectra of 
11.26~\mum\ \citep{slo07} to 11.29~\mum.

\cite{slo07} found that class C PAH spectra are associated 
with cooler radiation fields.  Massey SMC 59803 has an 
estimated effective temperature of 4100 K \citep{lev06}, and 
its PAH spectrum, while noisy, is consistent with this 
temperature.

\subsection{HV 5810} 

As discussed in \S 4.2 and illustrated in Figure 6, the 
spectrum of HV 5810 shows emission from several C$-$H bending 
modes in PAHs.  These include the in-plane bending mode at 
8.55 and the out-of-plane bending modes at 11.27, 12.78, and 
13.5~\mum\ in SL \citep[e.g.][]{atb89,hon01}.  The emission 
at 12.78~\mum\ probably contains a contribution from [Ne II] 
emission as well.  The LL portion of the spectrum in 
Figure 24 reveals an emission complex in LL with peaks at 
16.5 and 17.35~\mum, which also arises from PAHs 
\citep{mou00,wer04b}.

Examination of the spectral images reveals that the emission 
features at 11.3 and 12.8~\mum\ extend the length of the SL 
slit.  The emission at 8.6~\mum\ is also extended.  In the LL 
images, the 16--18~\mum\ PAH complex is extended over about 
$\sim$20$\arcsec$, while several forbidden emission lines are 
extended over much larger areas:  [\ion{Ne}{3}] at 15.6~\mum, 
[\ion{S}{3}] at 18.7 and 33.5~\mum, and [\ion{Si}{2}] at 
34.8~\mum.  The forbidden lines are commonly seen in many 
images of other targets, but they usually subtract out when 
the background is removed.  At longer wavelengths, the 
spatial structure of HV 5810 grows more complex and extended 
as well.

Figure 23 shows that HV 5810 is embedded in a complex
region of extended emission.  We conclude that we have 
observed a naked star, HV 5810, superimposed on a complicated
background.  

\subsection{IRAS 04530$-$6916} 

\begin{figure} 
\includegraphics[width=3.5in]{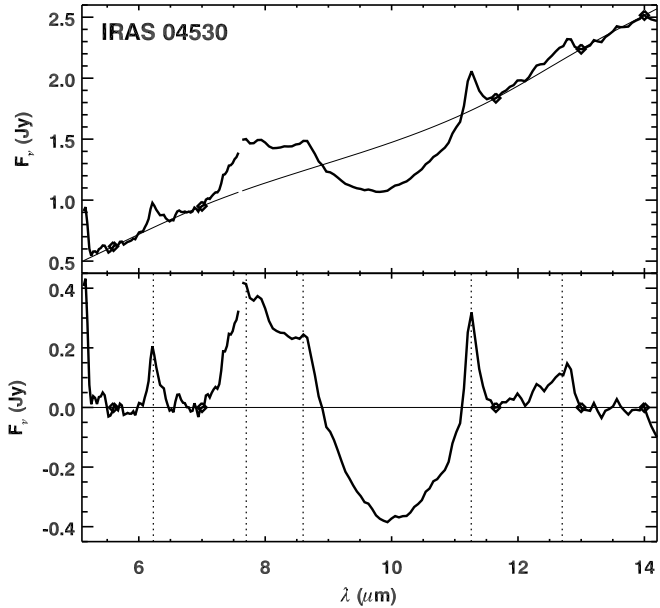}
\caption{The SL spectrum of IRAS 04530, scaled to match the
\iras\ photometry.  In the top panel, the spline points are
marked with diamonds, and the fitted spline with a smooth
thin curve.  The spline-subtracted spectrum appears in the
bottom panel, showing both silicate absorption and PAH
emission.  The vertical dotted lines are at wavelengths of
6.23, 7.70, 8.60, 11.26, and 12.70~\mum.}
\end{figure}

The observations of IRAS 04530 were partially mispointed in
both SL and LL, requiring some special attention to complete
the processing.  The source was mostly out of the slit in
one of the SL nod positions, and in LL, we only observed
a portion of the source, which is extended.  To stitch the
spectral segments together, we normalized them to the 12 and
25~\mum\ fluxes from the \iras\ Faint Source Catalog (2.01
and 4.97 Jy, respectively).  Because of the spectral tilts
in mispointed spectra, this leaves a discontinuity at the
SL/LL interface.  The overall continuum properties of the
resulting spectrum should be considered provisional.
Despite the limitations of the spectrum, the presence of
various spectral features and their wavelengths are reliable.

The spectrum shows a combination of silicate absorption
and emission from PAHs.  Although the IRS does not cover the 
35--45 \mum\ wavelength range that the Hanscom system used to 
distinguish the reddest sources, the \iras\ data show that 
the spectrum continues to rise between 25 and 60 \mum, making
this spectrum class 5.U/SA \citep[this class was
introduced by][]{ekp04}.

The combination of both PAH emission and silicate absorption 
complicates the analysis of the PAH strengths, but it does
not have a significant effect on their positions.  To examine
the PAH features more closely, we fit a spline through the
most likely continuum, as Figure 25 illustrates.  The
resulting PAH spectrum has features peaking at 6.22, 7.65,
8.62, 11.27, and 12.78~\mum.  With the exception of the last
feature, which may be affected by [\ion{Ne}{2}] emission, the
wavelengths for these features correspond to a class A PAH
spectrum in the scheme of \cite{pee02}.  This class of
PAH spectrum is commonly seen in star formation regions.


IRAS 04530 has usually been treated as an evolved star. 
\cite{woo92} measured its bolometric magnitude to be $-$7.62 
and concluded that it was probably a supergiant due to its 
high luminosity, low pulsation amplitude, and lack of maser 
activity.  \cite{vl01} seconded this opinion, and on the 
basis of these works it was included in this study.  
\cite{vl05a} recently suggested that it may instead be a YSO. 
Their optical spectrum shows strong hydrogen emission lines 
as well as the [\ion{S}{3}] doublet and [\ion{Ca}{2}] triplet. 
\cite{hod04} and \cite{ekp04} found that the U/SA 
classification in the \iso\ PHOT-S and CAM-CVF databases 
occurred almost exclusively in YSOs, star forming regions, 
and starburst galaxies (the single exception being C* 1662).

The SAGE images at 8~\mum\ and 24~\mum\ in Figure 23
reveal that IRAS 04530 lies at the edge of an extensive 
region of 8~\mum\ emission probably associated with the 
DEM L15 H II region.  The IRAC and MIPS photometry of 
IRAS 04530 \citep[from the 2006 December release of the 
SAGE point source catalog][]{mei06} and the 2MASS photometry 
help clarify its evolutionary status.  IRAS 04530
does not lie near the AGB population in the color-color
and color-magnitude diagrams published from Galactic and 
Magellanic Spitzer programs \citep{whi03a,whi03b,blu06,bol07}.

In summary, evidence from the new {\em Spitzer} observations, 
both spectroscopic and photometric, support the suggestion of 
\cite{vl05a} that IRAS 04530 is a YSO, not an evolved star.

\section{Dependencies on Metallicity} 

The previous sections concentrated on the analysis 
applied to the different classes of spectra and the results 
for individual sources.  This section concentrates on the 
samples as a whole, comparing the SMC and LMC samples to the 
Galactic samples published elsewhere.  The dominant difference 
between these samples is metallicity.

\subsection{Fraction of naked stars} 

\begin{deluxetable}{llll} 
\tablenum{9}
\tablecolumns{4}
\tablewidth{0pt}
\tablecaption{Fraction of naked oxygen-rich stars\label{Tbl8}}
\tablehead{ 
  \colhead{ } & \multicolumn{3}{c}{Period (days)} \\
  \colhead{Sample} & \colhead{$\le$250} & \colhead{250--700} & \colhead{$>$ 700}
}
\startdata
Galaxy & 2 of 106 & 5 of 269 & 0 of 11 \\
LMC    & 2 of 3   & 2 of 9   & 0 of 8 \\
SMC    & 1 of 1   & 6 of 7   & 0 of 1 \\
\enddata
\end{deluxetable}

Our sources were selected using criteria similar to those for
the Galactic samples of AGB stars examined by \cite{sp95} and 
supergiants studied by \cite{sp98}.  The Galactic samples were 
chosen by searching for optically identified variables of 
classes associated with either the AGB or supergiants in the 
LRS Atlas \citep{lrs86}.  \cite{sp95} found that 43\% of 
irregular (Lb) variables, 19\% of semi-regular (SRb) 
variables, and only 1\% of Mira variables (including SRa 
variables) were naked, while \cite{sp98} found that 13 of 65, 
or 20\%, of the supergiants were naked.  

The MC\_DUST sample includes five supergiants, all of which 
are associated with circumstellar dust, but the situation is 
different for the AGB.  In the LMC, three of the 19 oxygen-rich
sources, or 16\%, are naked (we are excluding the naked
star J052832, about which little is known), while in the SMC, 
seven of nine oxygen-rich AGB sources, or 78\%, are naked.

We know the pulsation periods of enough of the various 
samples to compare how the fraction of naked stars varies 
with period in the Galaxy, LMC, and SMC.  Because the number 
of naked stars is small, we divide the periods into only 
three bins separated at 250 d, the approximate boundary 
between most of the short-period semi-regulars and Miras, and 
700 d, the approximate period limit for Miras in the Galactic 
sample.  For the Galactic sample, we use the catalog 
published by \cite{sp98}.  Of the 611 oxygen-rich AGB stars 
and supergiants in that catalog, 386 have periods, and only 
seven of those are naked.  Two of the naked stars have 
periods less than 250 days, and the other five fall in the 
250--700-day bin, as Table 9 shows.  If we limited the sample 
to just Miras, the results are similar, with only two naked 
Miras out of 229.


Table 9 reveals that in the two lower-period bins, the
percentage of naked stars increases as the metallicity of
the sample decreases.  None of the stars
with periods greater than 700 days are naked.  In the SMC,
the only variable with a period less than 700 days and a
dust excess is BFM 1, but even its dust shell is thin, and
the dust spectrum is unusual.

\subsection{Dust Emission Contrast} 

\begin{figure} 
\includegraphics[width=3.5in]{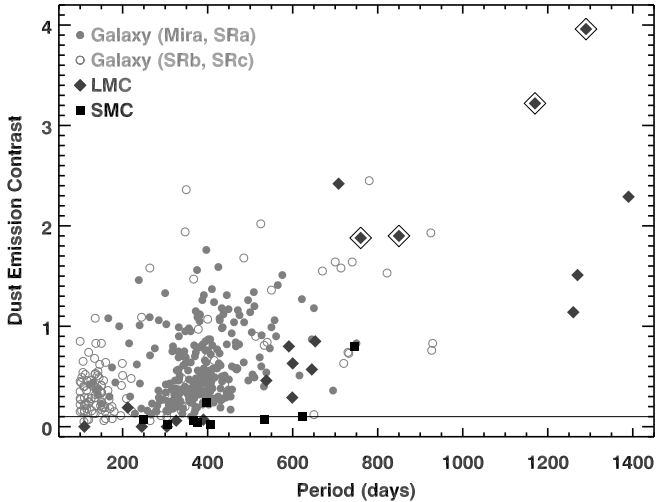}
\caption{Dust emission contrast among oxygen-rich souces as a
function of period, for all variables with known periods in 
the Galactic LRS sample \citep{sp95,sp98} and the MC\_DUST 
sample.  Dust emission contrast is defined in \S 3.4.  In the 
Galactic sample, Miras and the closely 
related SRa variables are plotted as filled circles, while 
the SRb and SRc variables are plotted as open circles.  LMC 
sources appear as dark gray diamonds, and SMC sources as 
black squares.  The four symbols outlined in black mark 
suspected massive stars.  The two points with periods 
$\sim$800 days are supergiants.  The two in the upper right 
are massive AGB sources (\S 3.3).  The horizontal line at 
DEC = 0.10 separates naked and dusty sources in the 
Magellanic samples.}
\end{figure}

\begin{deluxetable}{llll} 
\tablenum{10}
\tablecolumns{4}
\tablewidth{0pt}
\tablecaption{Dust emission contrasts\label{Tbl9}}
\tablehead{ 
  \colhead{ } & \multicolumn{3}{c}{Period (days)} \\
  \colhead{Sample} & \colhead{$\le$250} & \colhead{250--700} & \colhead{$>$ 700}
}
\startdata
Galaxy & 0.38$\pm$0.27    & 0.61$\pm$0.41 & 1.24$\pm$0.61 \\
LMC    & 0.06$\pm$0.11    & 0.44$\pm$0.30 & 2.29$\pm$0.92 \\
SMC    & 0.06$\pm$\nodata & 0.07$\pm$0.07 & 0.80$\pm$\nodata \\
\enddata
\end{deluxetable}

As part of the classification process in \S 5.1, we also 
measured the dust emission contrast for each oxygen-rich 
source, as defined for the naked stars in our sample.  
Figure 26 plots the dust emission contrast as a function of 
period in the Magellanic and Galactic samples for the naked 
and oxygen-rich sources.  For periods between 250 and 700 
days, the range dominated by Mira variables, the Galactic 
sample shows a general trend of increasing dust emission 
contrast with increasing period, but with a substantial 
spread.  For periods less than 700 days, the Magellanic 
sources show significantly less dust emission than their 
Galactic counterparts.  This result is consistent with the 
low SiO molecular band strength detected in naked oxygen-rich 
stars in the Magellanic Clouds, since SiO is the basic 
building block of silicate dust.


Magellanic sources with periods greater than 700 days 
typically show large contributions from dust in their 
spectra.  In particular, five sources lie on or appear to 
continue the trend of DEC vs.\ period established by the 
Galactic sample.  Two of these sources, HV 888 and HV 996,
are definite supergiants (at DEC$\sim$1.9 and P$\sim$800 d).
Two more, IRAS 04509 and IRAS 04512, are probably very 
luminous AGB sources as discussed in \S 3.3.
We believe these four sources are more massive than typical
AGB sources and therefore younger.  Since they formed more
recently, they are probably more metal rich than typical
Magellanic sources, and they would be expected to mimic the
more metal-rich Galactic sample.

Figure 26 includes three more LMC sources with very long
pulsation periods to the lower right.  From top to bottom,
these are IRAS 05402 (DEC$\sim$2.3), IRAS 04545 
(DEC$\sim$1.5), and IRAS 05329 (DEC$\sim$1.1).  All three
of these sources show silicate absorption at 10~\mum, so
our DEC measurement, which is a simple sum, underestimates 
the total silicate contribution.  Correcting for this 
simplification would undoubtedly push these points upward in 
Figure 26.  However, all three spectra show clear evidence for 
crystalline silicate emission at longer wavelengths, which
as described in \S 5.2, resemble the spectra from dusty
Galactic objects with disks.  This semblance is far from
direct evidence of a disk, but it might indicate that 
presence of a reservoir of dust in addition to the dust 
forming in the outflows from these stars at the current time.

Table 10 presents the mean dust emission contrasts for the 
same period ranges used for Table 9.  For the stars with 
periods $\le$700 days, the amount of dust drops from the 
Galaxy to the LMC and drops further in the SMC.  For these 
stars, the amount of dust clearly depends on metallicity.  
As described above, the samples with periods $>$ 700 days 
probably include young massive stars which blur the 
dependence on metallicity seen at shorter periods.


\subsection{SiO band strengths} 

\begin{deluxetable}{lr} 
\tablenum{11}
\tablecolumns{2}
\tablewidth{0pt}
\tablecaption{Comparative SiO band strengths\label{Tbl10}}
\tablehead{
  \colhead{Sample} & \colhead{W (\mum)} 
}
\startdata
SWS M giants & 0.35$\pm$0.11 \\
IRS M giants & 0.28$\pm$0.06 \\
LMC          & 0.26$\pm$0.11 \\
SMC          & 0.16$\pm$0.08 \\
\enddata
\end{deluxetable}

\cite{mat05} obtained 3--4~\mum\ spectra of six oxygen-rich 
evolved stars in the LMC, and they detected possible weak SiO 
absorption in only one source, IRAS 05218, which had been 
thought to be a carbon star \citep{tra99}.  They explained 
the absence of this band by noting that silicon is not 
produced in AGB stars, so its abundance in the photosphere 
reflects the initial abundance of the star, which will be low 
in a metal-poor system like the LMC.  But at least four of 
their six targets have circumstellar dust shells, which make 
it possible that the dust partially masks the SiO bands.

In the MC\_DUST sample, the SiO fundamental band is clearly
present in the spectra of several of the less obscured 
sources.  Table 11 compares the mean equivalent widths of the
total LMC and SMC samples to two Galactic samples, one 
based on SWS observations and a second based on IRS data.
For the SMC and LMC samples, we use only those equivalent 
widths in Table 6 with a S/N of $\sim$1 or better.  For 
sources with only an upper limit, we use it in place of the 
measured width.  For HV 2575 (first observation), we take the 
upper limit to be 0.26~\mum.  The follow-up spectra are 
treated as separate data.  MACHO 79.5505.26 is not included
since it may be carbon rich.


Using {\it ISO}/SWS data, \cite{her02} determined that the 
strength of the SiO fundamental band increases from an
equivalent width of $\sim$0.05~\mum\ in K0 giants to 
$\sim$0.18~\mum\ at K5.  No M giant had an equivalent width 
less than $\sim$0.15, while the maximum value increased past 
0.25~\mum\ at M5.  They measured the SiO equivalent width
between 7.6 and 9.0~\mum, but the bandhead is at 7.4~\mum,
and the band continues to 11.0~\mum\ in some cases.  We have
remeasured the SiO fundamental band in their sample of 23
M giants using a similar method to that applied to the IRS
data, except that we integrate the equivalent width directly
instead of fitting a template absorption band.  We shifted 
the long-wavelength range for fitting the Planck function 
from 11.5--13.0~\mum\ to 10.8--11.4~\mum\ to avoid hysteresis 
effects in the SWS data, which tend to drive the flux up 
slightly toward 12~\mum.  For similar reasons, we shifted the 
long-wavelength boundary for integrating the equivalent width 
from 11.5 to 11.1~\mum.  These shifts slightly reduce the 
equivalent widths, but the Galactic sample still has stronger 
SiO absorption than the Magellanic samples.  The mean SiO 
equivalent width shows no strong dependency on optical 
spectral type.

The second comparison sample of M giants comes from the
ongoing IRS GTO Cycle 4 program used to generate the SiO 
template described in \S 4.1.  At the time of this writing,
spectra of 13 of the 20 planned M giants have been obtained 
and reduced.  We extracted the SiO equivalent width using 
wavelengths identical to those used for the Magellanic 
samples, but as with the SWS data, we simply integrated the 
feature, instead of fitting a template.  As with the SWS
sample, the equivalent width does not depend strongly on 
spectral type.  The difference between the two Galactic M 
giant samples may result from different selection criteria,
but it is within the standard deviations.

Table 11 shows a general trend toward decreasing SiO
band strength with decreasing metallicity, although there
is some overlap in the samples.  The LMC sample overlaps
the IRS M giants.  The wide spread in the LMC sample may 
indicate a range of metallicities, but the sample is not 
large.  The SMC clearly shows weaker SiO bands than the 
Galactic giants.  These results confirm the 
metallicity-dependent trend first recognized by 
\cite{mat05} in the 3~\mum\ spectral region.

\subsection{Oxygen-rich dust composition} 

\begin{figure} 
\includegraphics[width=3.5in]{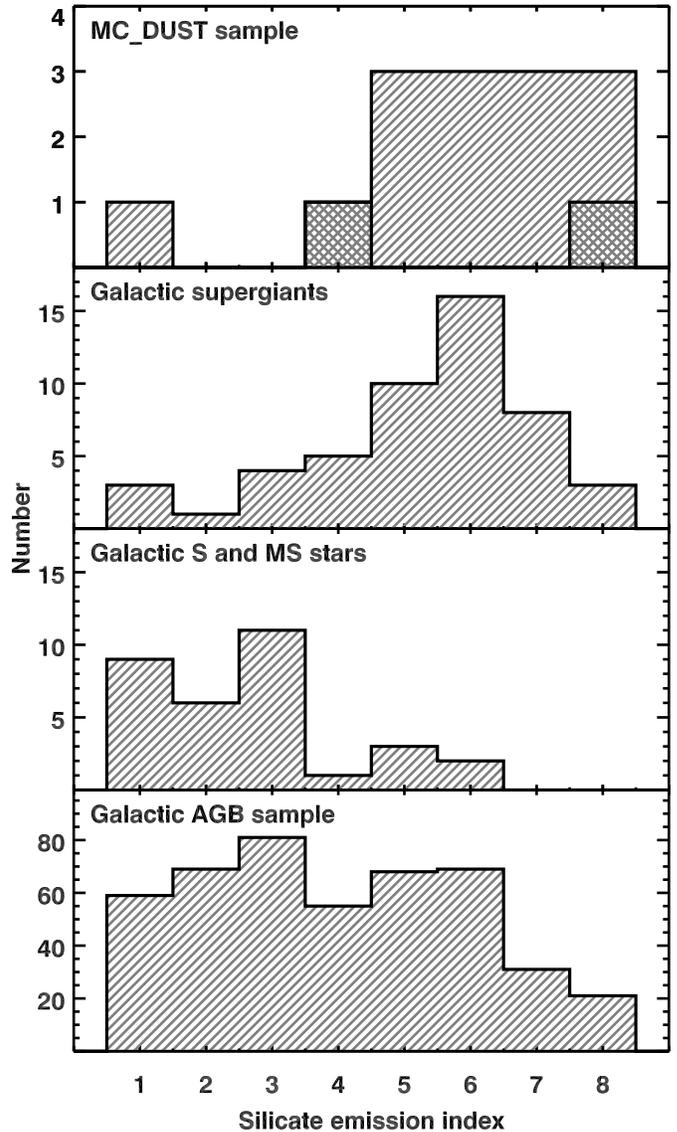}
\caption{The distribution of the MC\_DUST sample and
Galactic comparison samples with silicate emission index.  In 
the top panel, LMC sources are shaded with diagonal stripes, 
while the two SMC sources are cross-hatched (at SE4 and SE8).
Alumina-rich dust produces the SE1--3 spectra and is largely
absent from the Magellanic samples.  The dust distribution in
the Magellanic sample closely resembles the Galactic 
supergiants, even though only about three of the 14 
Magellanic sources plotted in the upper panel are actually 
supergiants.}
\end{figure}

\cite{es01} demonstrated with radiative transfer models that
the amorphous alumina grains produce the broad, low-contrast
features peaking $\sim$11--12~\mum, which correspond to lower
SE indices (SE1--3), while amorphous silicate grains produce
the classic 10~\mum\ silicate feature (and associated 
18~\mum\ feature) which corresponds to higher SE indices 
(SE6--8).  The middle portion of the sequence (SE4--6) 
generally shows structured emission features, with a peak at 
10~\mum\ and a shoulder at 11~\mum.  This structure could 
arise from multiple causes, including mixtures of amorphous 
alumina and silicates, self-absorbed silicates, or possibly, 
crystalline silicate grains.  Given that alumina and 
silicates both contribute to the spectra, it might be more
appropriate to refer to the sequence as the oxygen-rich dust 
sequence.

Figure 27 compares the distribution of silicate emission 
sources in the Magellanic Clouds to the Galactic samples 
examined by \cite{sp95,sp98}.  The Magellanic sample differs 
markedly from the Galactic AGB sample by having only two
sources from SE1 to SE4.  In fact, it looks more like the 
Galactic supergiant sample than anything else, which is 
somewhat surpising given that the sample displayed in 
Figure 27 includes only three supergiants.  We had 
intentionally observed ten stars classified as S or MS, 
which we had expected to resemble the Galactic S and MS 
stars, but most of these proved to be naked instead.  Of 
the five S and MS stars in the LMC, three (WBP 104, 
HV 2575, and HV 12620) are naked, while HV 12070 is an 
SE6 and WBP 77 accounts for the lone SE1 source.  Of the 
five in the SMC, only BFM 1 shows any sign of dust, but as 
discussed in \S 5.6., it is very unusual dust.

The lack of emission from amorphous alumina dust in the 
Magellanic Clouds is consistent with how the abundance of Al 
depends on metallicity compared to other dust-producing 
elements like Mg and Si.  The latter elements are both 
$\alpha$-capture products, and while their abundance is
lower in metal-poor stars, Al is even less abundant
\citep[see][and references therein]{whe89}. 

The apparent absence of a 13~\mum\ feature in our sample
provides further evidence of the relative lack of Al in the 
Magellanic Clouds.  The carrier of the 13~\mum\ feature has 
been the subject of some debate, with the focus on crystalline 
alumina \citep[corundum, Al$_2$O$_3$][]{gla94,slo96} or spinel 
\citep[MgAl$_2$O$_4$][]{po99,fab01a}.  \cite{slo03} and
\cite{dep06} have provided strong evidence against spinel 
as the carrier, but here, it is sufficient to note that in
both proposed carriers, the Al--O bond is responsible for
the 13~\mum\ feature.  Consequently, the lack of Al 
in the Magellanic Clouds fully explains its absence.

The presence of the 14~\mum\ feature in our Magellanic
sample provides some clues about its carrier.
\cite{pit06} recently noted that the 14~\mum\ feature in 
HV 2310 could be reproduced by AlN dust or possibly by 
non-spherical corundum grains.  While the non-spherical 
grain models discussed by \cite{dep06} do shift the emission 
peak from 13~\mum\ to longer wavelengths, the features become 
quite broad compared to the observed 14~\mum\ features.  In 
addition, the low abundance of Al in the Magellanic Clouds
make any Al-bearing dust species an unlikely carrier.
The presence of the 14~\mum\ feature appears to be limited to 
strong silicate emission sources, which suggests that its 
carrier is probably related to silicates in some way.



\subsection{Carbon mass-loss rates} 

\begin{figure} 
\includegraphics[width=3.5in]{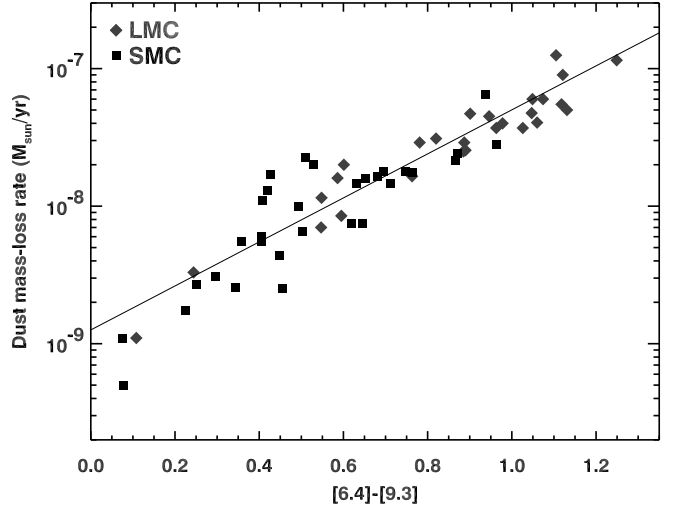}
\caption{Dust mass-loss rate as a function of [6.4]$-$[9.3]
color, including data from the samples of Magellanic carbon 
stars presented by \cite{slo06b,zij06} and \cite{lag07} and 
analyzed with radiative transfer models by \cite{gro07}.  We 
have converted the total mass-loss rates to dust mass-loss 
rates by dividing by their assumed gas-to-dust ratio of 200.  
The solid line is a linear fit to the data.}
\end{figure}

\begin{figure} 
\includegraphics[width=3.5in]{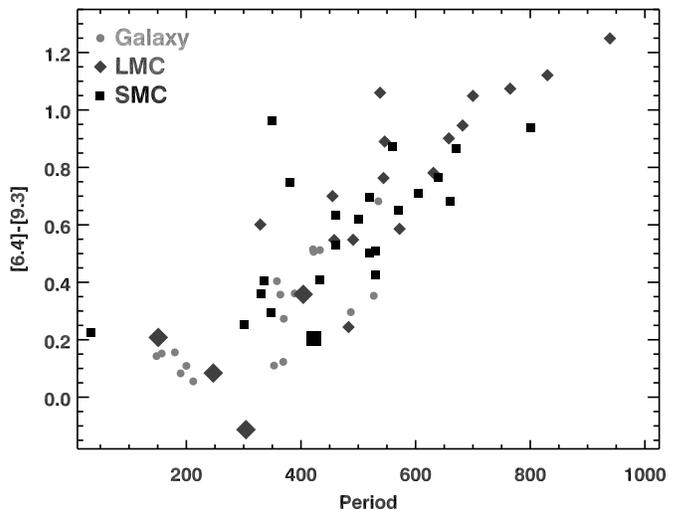}
\caption{[6.4]$-$[9.3] color as a function of period.  
The data include the samples from Fig. 22 (small diamonds
and squares), the MC\_DUST sample (large diamonds and
squares), and the Galactic sample (small circles).
The Galaxy, SMC, and LMC show no significant differences
despite their different metallicities.}
\end{figure}

Grains composed primarily of carbon atoms, usually described
as amorphous carbon, dominate the the dust produced by carbon 
stars \citep[e.g.][]{mr87,se95}, but amorphous carbon has no 
easily discernable spectral features, making a direct 
measurement of its contribution difficult.  The Manchester 
method substitutes the [6.4]$-$[9.3] color, which will grow
increasingly red as the contribution of amorphous carbon
relative to the star increases.  \cite{gro07} applied 
radiative transfer models to all of the Magellanic carbon 
stars in the previously cited papers, and they found a good 
correlation between total mass-loss rate and [6.4]$-$[9.3]
color (see their Fig. 7).  What the radiative transfer models 
actually probe, however, is the {\it dust} mass-loss rate.  
The total mass-loss rate is simply inferred from the 
gas-to-dust ratio, which \cite{gro07} assumed to be 200 in
both the LMC and SMC.  We convert their mass-loss rates
to dust mass-loss rates (by dividing by 200) and repeat their
comparison to [6.4]$-$[9.3] color in Figure 28.  Both the LMC 
and SMC follow the same relation.  Fitting a line to the data 
gives log (mass-loss rate) $= -9.8 + 1.6 $ ([6.4]$-$[9.3]).  
The scatter of the data about this line has a standard 
deviation of 16\%.


This and previous studies have compared objects in the 
Galaxy, SMC, and LMC at similar [6.4]$-$[9.3] color as though
objects with similar amounts of amorphous carbon dust are 
truly alike.  They might not be if the dust formation 
efficiency depended on metallicity.  To examine this 
possibility, we examined the pulsation periods of the sources 
in the various samples, since those should be similar for 
physically similar stars.  Figure 29 plots [6.4]$-$[9.3]
color as a function of pulsation period, and it reveals no
significant differences in the distribution of colors at a 
given period between the samples.  In other words, carbon 
stars with similar pulsation properties are embedded in 
similar quantities of dust, regardless of their initial 
metallicity.


\section{Discussion and Conclusions} 

The similar dust production rates among the samples in
Figure 29 have important consequences.  The production of 
amorphous carbon is limited by the amount of excess carbon 
compared to oxygen, but carbon is produced in the 
triple-$\alpha$ process in AGB stars.  The production of 
silicates and other oxygen-rich grains, on the other hand, 
is limited by the abundances of heavy elements such as Si 
and Al, and these are not produced in significant quantities 
on the AGB.  The ready explanation for the results 
presented in Figure 29 is that the carbon produced and 
dredged up by AGB stars dominates the carbon with which they 
formed.  It should follow, then, that even Population III 
stars, if they are in the mass range to become carbon stars, 
will produce approximately the same amount of carbon-rich 
dust as their Population I counterparts.

\cite{bbm78,bbm80} discovered that in the SMC, more AGB
stars became carbon rich than in the LMC \citep[see 
also][]{ch03}.  \cite{rv81} explained this phenomenon as a 
metallicity effect.  Stars with solar abundances will become 
carbon stars if their initial mass $\ga$1.7~M$_{\sun}$, but
at 1/5 solar metallicity, the mass limit has dropped to 
1.2~M$_{\sun}$, primarily because of the lack of oxygen in
the atmosphere of the metal-poor star to balance the dredged
up carbon.  Assuming a Salpeter initial mass function 
\citep{sal55} and integrating from these carbon limits to
8~M${_\sun}$, we find that in the SMC, 70\% more of the
AGB should be carbon-rich compared to a sample with solar
metallicity.  Since we see no diminishment of carbon-rich 
dust production at decreasing metallicity, we conclude that 
the more metal-poor an initial population of stars, the more 
carbon-rich dust it will contribute to the ISM, simply 
because it produces more carbon stars.  

Many other properties of carbon stars do depend on 
metallicity.  Our small sample reinforces results from
previous studies of {\it Spitzer} spectroscopy in the 
Magellanic Clouds, which show that as the metallicity
decreases, emission from SiC and MgS dust decrease, 
while absorption from acetylene gas increases.  The
increased acetylene absorption probably arises from 
higher C/O ratios in Magellanic carbon stars 
\citep{vl99,mat02,mat05}, which in turn arise from a 
combination of increased dredge-up efficiency \citep{woo81} 
and less initial oxygen in the atmosphere \citep{vw93}.
The additional acetylene we observe may add even more to
the amount of carbon injected into the ISM by metal-poor
populations.

The behavior of the oxygen-rich dust stands in stark contrast
to the carbon-rich dust.  While the quantity of carbon-rich
dust shows no discernable dependence on metallicity, 
oxygen-rich dust shows many clear trends.  As metallicity
decreases, (1) the fraction of naked stars increases, (2)
SiO absorption decreases, (3) the silicate dust emission also 
decreases, and (4) alumina-rich dust emission disappears 
almost entirely.  The lower metallicity of the Magellanic
Clouds provides a ready explanation for all of these effects.
Si is not produced on the AGB, so its abundance scales with 
overall metallicity.  As a result, we observe less SiO gas,
which is the fundamental building block of silicate dust, 
and less silicate dust as well.  The abundance of Al scales
even more strongly with metallicity (\S 5.4.), so the
lack of alumina dust should not be a surprise.  Not only 
do we see few spectra in the SE1--3 range, but we see little
or no emission from the 13~\mum\ feature, which arises from
an Al--O bond.

Standard condensation sequences propose the formation of
alumina-rich grains before silicates 
\citep[e.g.][]{tie90,ste90}.  The lack of alumina dust in 
our sample suggests the formation of alumina is not a
prerequisite to the formation of silicates.  In other words,
silicates can form without pre-existing alumina grains to 
serve as seeds.

The oxygen-rich sample contains a wealth of interesting dust 
features.  Five sources show crystalline silicate features 
at 10~\mum\ or longer wavelengths, and these are all 
consistent with Mg-rich pyroxene and olivine.  There is no 
evidence for Fe in the silicates, much like the Galactic 
sample.  Several sources have an unidentified 14~\mum\ 
emission feature, and one source, the only true S star in 
our sample, shows a previously unknown 13.5~\mum\ emission 
feature.  These two features are newly discovered by the
IRS, and their identification awaits further investigation.

The Magellanic Clouds present us with a scenario of steady or 
increasing carbon-rich dust production as systems grow more 
metal-poor, compared to declining rates of oxygen-rich dust 
production.  This result measures what goes into the ISM, but 
it contradicts observations of what is currently present in 
the ISM in metal-poor systems.  Most lines of sight in the 
SMC have unusual extinction curves without the commonly 
observed 2175~\AA\ bump seen in the Galaxy, in common with 
starburst galaxies and quasars \citep[e.g.][and references 
therein]{gc98,pit00}.  The 2175~\AA\ bump is commonly 
associated with graphitic carbonaceous material \citep[as 
first proposed by][]{sd65}.  \cite{ds98} showed that certain 
PAHs and PAH clusters can reproduce the 2175~\AA\ bump and 
that these particles can result from interstellar processing 
of hydrogenated amorphous carbon.  The recent study of 
unusual PAH-like spectra by \cite{slo07} demonstrates that 
hydrocarbons can exist with a range of aromatic-to-aliphatic 
ratios.  Many of the sources of unusual PAH-like emission are 
carbon-rich post-AGB objects.  All of these observations 
support an evolutionary relationship from amorphous carbon 
produced by carbon stars on the AGB and to interstellar PAHs 
to the more graphitic material responsible for the 2175~\AA\ 
bump.  The results in this paper demonstrate that carbon-rich 
dust production does not decrease in more metal-poor samples, 
and yet, metal-poor systems like the SMC are missing the 
2175~\AA\ bump that should result.

Either the missing carbon-rich dust did not form, or it has
been destroyed.  \cite{gc98} suggested that harsher radiation 
fields in metal-poor systems can destroy the carrier of the 
2175~\AA\ bump.  Stellar modelling shows that decreasing the 
metallicity of a star results in less UV line blanketing and 
an emitted spectrum with more UV emission \citep{giv02}, which 
readily explains the harsher interstellar radiation observed 
in metal-poor \ion{H}{2} regions \citep{mh02}.  \cite{gal08} 
recently reached the opposite conclusion.  They used galactic 
evolution models to investigate the absence of PAH emission 
in blue compact dwarf galaxies and other metal-poor galaxies 
\citep{eng05,wu06}, and concluded that these systems produce 
significantly less carbon-rich dust than the Galaxy.  Our 
sample of Magellanic carbon stars provides no evidence for 
less carbon-rich dust formation in more primordial systems, 
which leads to the conclusion that the lack of carbon-bearing
dust in these systems results not from reduced production,
but from enhanced destruction.

\acknowledgements

The authors thank the anonymous referee for carefully reading
the manuscript and helping us to improve it.
These observations were made with the {\it Spitzer Space
Telescope}, which is operated by JPL, California Institute of
Technology under NASA contract 1407 and supported by NASA
through JPL (contract number 1257184).  This research has
made use of the SIMBAD and VIZIER databases, operated at the
Centre de Donn\'{e}es astronomiques de Strasbourg, and the
Infrared Science Archive at the Infrared Processing and
Analysis Center, which is operated by JPL.

\clearpage

\clearpage
\thispagestyle{empty}
\setlength{\voffset}{15mm}
\clearpage
\setlength{\voffset}{0mm}

\clearpage

\clearpage

\end{document}